\documentclass[article,superscriptaddress]{revtex4-2} 
\UseRawInputEncoding
\usepackage{graphicx}
\usepackage{subfig}
\usepackage{color}
\usepackage{epstopdf}
\usepackage{amsfonts}
\usepackage{amsmath}
\begin{document}

\title{Weber number and the outcome of binary collisions between quantum droplets}

\author{ J. E. Alba-Arroyo}
\author{ S. F. Caballero-Benitez}
\author{R. J\'auregui}  
 \email{rocio@fisica.unam.mx}
\affiliation{%
Departamento de F\'{\i}sica Cu\'antica y Fot\'onica\\ Instituto de F\'{\i}sica, Universidad Nacional Aut\'onoma de 
M\'exico\\ Cd. de M\'exico C.P. 04510,
M\'exico
}

\date{\today}
\begin{abstract}
A theoretical analysis of  binary collisions of quantum droplets under feasible experimental conditions  is reported. Droplets formed from  degenerate dilute Bose gases made up from binary mixtures of ultracold atoms are considered. Reliable expressions  for  the  surface tension of the droplets are introduced based on a study of low energy excitations of their ground state within the random phase approximation. Their relevance is evaluated considering an estimation of the expected excitation energy having in mind the Thouless variational theorem. The surface tension expressions allow calculating the Weber number of the droplets involved in the collisions. Several regimes on the outcomes of the binary frontal collisions that range from  the coalescence of the quantum droplets to their disintegration into smaller droplets are identified. Atoms losses of the droplets derived from self- evaporation and three--body scattering are quantified for both homo- and hetero-nuclear mixtures. Their control is mandatory for the observation of some interesting effects arising from droplets collisions.
\end{abstract}

\maketitle

\section{Introduction.}

In most circumstances, degenerate states of dilute atomic gases are realized using  samples trapped by different means. The question of whether they can also exist in isolation, without a trap potential, was  theoretically  answered in the affirmative within two scenarios. The first one corresponds to a homogeneous atomic gas where the two-body scattering length is negative and much larger than the effective two-body interaction radius; then the contribution to the ground state energy due to the three--body correlations could  become dominant and provide a self--trapping mechanism  \cite{Bulgac2002}.
The second scenario corresponds to a condensed Bose-Bose mixture where the interspecies attraction becomes stronger than the geometrical average of the intra--species repulsions \cite{Petrov2015}. The stabilization mechanism is then provided by quantum fluctuations that are a direct manifestation of beyond mean-field effects. The implementation of the latter mechanism was the basis for the  experimental observation of self-bound droplets of ultra cold atoms in homo--nuclear~\cite{Cabrera2018,Semeghini2018} and hetero--nuclear\cite{DErrico2019} mixtures of alkali atoms, and in dipolar condensates \cite{Ferrier2016,Schmitt2016,Chomaz2016,Tanzi2019}. In the latter case, it was required to drive a Bose-Einstein condensate (BEC) into the strongly interacting regime.

The first correction to the mean-field contribution to the ground state energy of a homogeneous weakly repulsive Bose gas is given by the Lee-Huang-Yang (LHY) term~\cite{LHY1957}. This term is negligible in many circumstances. However,  there are situations where the LHY and the mean field contributions can be of the same order without  leaving the weakly interacting regime.  For a Bose-Bose mixture where the inter-- and intra--species coupling constants  are independently controlled, the mean--field term, which is proportional to the square of the density $n^2$, could be negative and the LHY contribution, which is proportional to $n^{5/2}$, could be positive. Then, the quantum LHY repulsion could neutralize the mean-field attraction and stabilize the system against collapse~\cite{Petrov2015}. For a large enough number of atoms, the resulting self--trapping quantum state is characterized by a core with a quasi uniform density and a thin free surface similar to that of a liquid droplet. A less intuitive, though more precise description of the self--trapping mechanism can be obtained via Quantum Monte Carlo simulations~\cite{bottcher2019}, they predict similar general properties for the quantum droplets, though different specific scenarios of their occurrence. The general properties are:
 generic features of the collective modes~\cite{Chomaz2016}, the presence of striped states in confined geometries~\cite{wenzel2017}, and the existence of arrays of phase coherent droplets with transient super solid properties~\cite{botcher2019,Tanzi2019,chomaz2019}. 

 For small atom numbers, the quantum droplet has no collective modes with energy lower than the particle emission threshold so that self-evaporation can occur~\cite{Petrov2015,Ferioli2020}. In this work we consider droplets with small and large atom numbers. For the latter, self-evaporation is not expected and the resulting  quantum state is closer to the liquid droplet expectations, {\it i.e.}, it exhibits properties nearby a classical liquid, a helium nano droplet, or an atomic nuclei within the liquid drop model.  Note, however, that in this regime three--body recombination collisions may reduce considerably the lifetime of the quantum droplets created by a binary mixture of Bose-Bose ultra cold atoms. Thus, we take into consideration  recent experimental results~\cite{DErrico2019} which demonstrate that, for an adequate choosing of hetero-nuclear mixtures,  quantum droplets with lifetimes in the order of tens of milliseconds can be produced.

The collisions between droplets is an interesting mechanism both to study the quantum droplet excitations, and to realize quantum simulations of analog many body systems. Up to now, this mechanism has been used to observe the crossover between compressible
and incompressible regimes of quantum droplets~\cite{Ferioli2019}. In the incompressible regime surface waves occur and surface tension should be a relevant parameter to understand the conditions required for different results of quantum droplet collisions.

Our analisis is structured as follows. Following the ideas implemented by the nuclear physics community to study the expectations of the nuclear  liquid drop model~\cite{berstch1974,berstch1977, casas1990}, expressions for the low energy surface excitations are used as a variational {\it ansatz}. These expressions include effects directly due to the thickness of the surface of the droplet, and allow to quantify the surface tension of the ground state droplet. We  evaluate numerically  -- within the extended Gross-Pitaevskii equation (EGPE) that incorporates the LHY terms  in three dimensions-- the ground state for a finite number of atoms, ranging from $10^4-10^7$ in Bose-Bose mixtures. The analysis is performed for binary mixtures of homo- and hetero--nuclear atomic gases within the conditions on the interaction strengths that give rise to quantum droplets. In the case of homo--nuclear BEC mixtures, we also obtain the ground state  within an effective formalism that considers finite-range effects as developed in Refs.~\cite{bottcher2019,Cikojevic2018}.  General features of the resulting ground states are then evaluated. They include the  surface tension derived from the different surface excitation models described in the previous Section, and the dependence of the droplets radius and saturation energy with the number of atoms in the droplets. Whenever it is possible, the results are compared to previous theoretical predictions.
In the next Section, atom losses due to self-evaporation and three--body recombination effects are analyzed. Finally,  the outcomes of frontal binary collisions are  explored. For classical liquid droplets the collisions are usually investigated in terms of the impact parameter and of the Weber number. The latter is a measure of the relative importance of the inertia of a fluid in terms of the kinetic energy compared to its surface tension~\cite{frohn2000}. In this work, several regimes for binary front collisions that range from  the coalescence of the quantum droplets to their disintegration into smaller droplets are identified in terms of the Weber number.

\section{Low energy exciations of a liquid-like system.}
We are interested in obtaining expressions for the surface tension of a quantum droplet. To that end we shall follow the formalism developed by Berstch in the study of capillary waves in a superfluid~\cite{berstch1974}, and which was later extended to Fermi systems with spherical symmetry~\cite{berstch1977} having in mind the liquid drop model of nuclei.
This formalism is based on  the random phase approximation (RPA) and the Thouless variational theorem.
In our case, the random phase approximation is equivalent to  the linearization of the time-dependent equations that define the dynamics of the system in the vicinity of a variational minimum; Thouless variational theorem guarantees that  actual excitation energies are a lower bound to the excitation energies that result from using, as an {\it ansatz}, known expressions of excitations with harmonic time dependence and satisfying adequate boundary conditions.

Let us consider a binary mixture of Bose gases composed by $N^{(a)}$ and $N^{(b)}$ atoms of each species. Let us assume that the  state of the system is approximately described by the Hartree many-body wavefunction 
\begin{equation}
\Psi_N(\vec r_1,...,\vec r_{N^{(a)}};\vec r^\prime_1,...,\vec r^\prime_{N^{(b)}}) = \Big[\prod_{i=1}^{N^{(a)}}\chi^{(a)}(\vec r_i)\Big]\Big[\prod_{i=1}^{N^{(b)}}\chi^{(b)}(\vec r^\prime_i)\Big].
\end{equation}
As  order parameters we consider
\begin{equation}
\psi^{(\alpha)}(\vec r,t) = \sqrt{N^{(\alpha)}} \chi^{(\alpha)}(\vec r,t),\quad\quad  \alpha = a, b,\label{eq:normalizacion}
\end{equation}
$\psi^{(\alpha)}$ evolves according to the equation
\begin{eqnarray}
i\hbar \partial_t \psi^{(\alpha)}(\vec r,t) &=& \hat H\psi^{(\alpha)}(\vec r,t)\\
&=& \hat H_0^\alpha \psi^{(\alpha)} + \Big(\int dr^3 dr^{\prime 3} U_{\alpha\alpha}(\vec r, \vec r^\prime;\rho^{(\alpha)}(\vec r^\prime,t)) + U_{\alpha\beta}(\vec r, \vec r^\prime;\rho^{(\beta)}(\vec r^\prime,t))\Big)\psi^{(\alpha)}(\vec r,t), \quad \alpha\ne \beta\label{eq:evol}\\
\hat H_0^\alpha &=& -\frac{\hbar^2}{2m_\alpha}\nabla^2 + V^\alpha_{ext}(\vec r)\label{eq:H0}
\end{eqnarray}
In Eq.~(\ref{eq:evol}), $V^\alpha_{ext}(\vec r)$ represents an external potential and $\rho^{(\alpha)}=\psi^{(\alpha)*}\psi^{(\alpha)}$ the density of  $\alpha$--atoms. 
The potentials $U_{\alpha\beta}$ may incorporate, in an effective way, interactions beyond the standard mean field approximation. In the case of the extended Gross-Pitaeviskii formalism, the density dependent interactions are superpositions of contact terms, and have the structure,
\begin{equation}\label{eq:EGPEE}
 U_{\alpha\beta}(\vec r, \vec r^\prime;\rho^{(\gamma)}(\vec r^\prime,t))  = \delta(\vec r -\vec r^\prime) \sum_i g^i_{\alpha\beta}(\rho^{(\gamma)}(\vec r, t))^{s_i},\quad \quad \alpha,\beta,\gamma =a,b,
\end{equation}
with $s_i$ a real number.
The stationary solutions of Eq.~(\ref{eq:evol}) define the chemical potentials $\mu_\alpha$,
\begin{equation}
\psi^{(\alpha)}_0 (\vec r,t)= e^{-i\mu_\alpha t/\hbar}\phi^{(\alpha)}_0(\vec r),\quad\quad  \alpha = a, b.\label{eq:timedependent}
\end{equation}
Let us consider collective Bogolubov excitations with a harmonic time dependence,
\begin{equation}
\psi^{(\alpha)}_{exc} (\vec r,t)= e^{-i\mu_\alpha t/\hbar}\Big(\phi^{(\alpha)}_0(\vec r) + \sqrt{N^{(\alpha)}}
\sum_q (u_q^{(\alpha)}(\vec r) e^{-i\omega_q t} + v_q^{(\alpha)*}(\vec r) e^{-i\omega_q t}) 
\Big).
\end{equation}
In this equation $q$ is a label that determine the characteristics of the excitation derived, for example, from its geometry, excitation energy etc.
The  excitation functions belong to the space expanded by a normalized and complete basis set $\{u_q^a,v_q^a,u_q^b,u_q^b\}$, according to the scalar product
\begin{equation}
\int d^3r [u_q^{(\alpha)}(\vec r)u_p^{(\alpha) *}(\vec r) -v_q^{(\alpha)}(\vec r)v_p^{(\alpha) *}(\vec r)] = \frac{\omega_q}{\vert \omega_q\vert} \delta_{qp}, \quad \omega_q\ne 0.\label{eq:nor}
\end{equation}
Excitations with $\omega_q = 0$ are assumed to be orthogonal to $\phi^{(\alpha)}_0$.
Within the linear response approximation around the stationary function $\psi^{(\alpha)}_0 (\vec r,t)$ , Eq.~(\ref{eq:evol}) is equivalent to linear equations with the structure
\begin{equation}
{\mathcal{M}}
\begin{pmatrix}u_q^{(a)}\\v_q^{(a)}\\u_q^{(b)}\\v_q^{(b)}\end{pmatrix} = 
\hbar\omega_q 
\begin{pmatrix}u_q^{(a)}\\-v_q^{(a)}\\u_q^{(b)}\\-v_q^{(b)}\end{pmatrix}.
\end{equation}
Thouless variational theorem establishes that the energy $\hbar \omega_{exact}$ of harmonic exact solutions of Eq.~(\ref{eq:evol}) are a lower bound of $\hbar \omega_q$. This condition is written in our case as
\begin{equation}
\hbar \omega_{exact}\le \frac{1}{2}\frac{\sum_{\alpha,\beta=a,b} \langle \eta_q^{(\alpha)}\vert \hat \Xi^{\alpha\beta}\vert \eta_q^{(\beta)}\rangle + \sum_{\alpha = a,b}\langle \zeta_q^{(\alpha)}\vert \hat \Delta^\alpha\vert \zeta_q^{(\alpha)}\rangle}{\sum_{\alpha=a,b} (\langle \zeta_q^{(\alpha)}\vert \eta_q^{(\alpha)}\rangle)}=\hbar \omega_{\eta\zeta}
\end{equation}
where
\begin{eqnarray}
\eta_q^{(\alpha)} &=& \sqrt{N^{(\alpha)}}(u_q^{(\alpha)} + v_q^{(\alpha)*}),\quad \zeta_q^{(\alpha)} =\sqrt{N^{(\alpha)}}(u_q^{(\alpha)} - v_q^{(\alpha)*}),\\
\hat\Delta^\alpha &=& \hat H_0^\alpha + \int dr^3 \int dr^{\prime 3}\big(U_{\alpha\alpha}(\vec r, \vec r^\prime;\rho^{(\alpha)}_0(\vec r^\prime)) + U_{\alpha\beta}(\vec r, \vec r^\prime;\rho^{(\beta)}_0(\vec r^\prime)\big) - \mu_\alpha,\\
\hat\Xi^{\alpha\beta}\eta_q^{(\beta)}(\vec r) &=& \delta_{\alpha\beta}\hat \Delta^\alpha \eta_q^{(\beta)}(\vec r) +
2\phi^{(\alpha)}_0(\vec r)\int d^3r^{\prime}  U_{\alpha\beta}(\vec r, \vec r^\prime;\rho^{(\beta)}_0)\phi^{(\beta)}_0(\vec r^\prime)\eta_q^{(\beta)}(\vec r^\prime).
\end{eqnarray}
Our {\it ansatz} for the excitation functions depend functionally on $\phi_0$ and its derivatives
\begin{equation}
\eta_q^{(\alpha)} = \eta_q^{(\alpha)}[\phi_0^{(\alpha)},\partial \phi_0^{(\alpha)}]
,\quad
\zeta_q^{(\alpha)} = \gamma_\alpha \tilde{\zeta}_q^{(\alpha)}[\phi_0^{(\alpha)},\partial \phi_0^{(\alpha)}].
\end{equation}
The $\gamma^\alpha$  parameters are determined  by requiring to yield a minimum for $\omega_{\eta\zeta}$, the result is
\begin{eqnarray}
\gamma_a &=& C_a\sqrt{\frac{AB_b}{B_a(B_b C_a^2 + B_aC_b^2)}},\quad \gamma_b =\frac{B_aC_b}{B_bC_a}\gamma_a\\
 A &=& \sum_{\alpha,\beta=a,b} \langle \eta_q^{(\alpha)}\vert\hat\Xi^{\alpha\beta}\vert \eta_q^{(\beta)}\rangle,\quad
 B_\alpha= \langle\tilde{\zeta}^{(\alpha)}_q\vert \hat\Delta^\alpha   \vert\tilde{\zeta}^{(\alpha)}_q\rangle,\quad
C_\alpha =\langle \tilde{\zeta}_q^\alpha\vert \eta_q^\alpha\rangle.
\end{eqnarray}
The optimal variational excitation energy satisfies the equation
\begin{equation}\label{eq:omegaopt}
\hbar^2\omega_{opt}^2 = \frac{AB_aB_b}{B_aC_b^2 + B_bC_a^2}.
\end{equation}
We shall consider functions $\eta$ and $\tilde{\zeta}$ with the simple structure,
\begin{equation}\label{eq:ansatz}
\eta^\alpha_q = \vec F_q^\alpha(\vec r)\cdot\xi^{(\alpha)}\vec\nabla\phi_0^{(\alpha)}, \quad \tilde{\zeta}^\alpha_q = \Lambda^\alpha_q(\vec r)\phi_0^{(\alpha)}. 
\end{equation}
The functions $\vec F^\alpha(\vec r)$ and $ \Lambda^\alpha (\vec r)$ are usually chosen to incorporate geometric properties of the droplets surface. The parameter $\xi^{(\alpha)}$  corresponds to a characteristic scale length for the $\alpha$-fluid. 
 A direct calculation shows that, in case  the dynamics is governed by an extended Gross Pitaevskii equation characterized by contact interactions, Eq.(\ref{eq:EGPEE}), the contribution of the interaction terms of the excitation functions Eq.~(\ref{eq:ansatz}) to $A$ and $B_\alpha$ is null. However, information about such interaction terms is still contained in the stationary solution $\phi_0^{(\alpha)}$.

The particular selection
\begin{equation}
\vec F_q^\alpha(\vec r) =\xi^{(\alpha)}\nabla  \Lambda_q^\alpha(\vec r),\label{eq:Fq}
\end{equation}
allows a hydrodynamic interpretation of $\zeta^{(\alpha)}_q/\phi_0^{(\alpha)}$ as a velocity field asociated to an infinitesimal change of the density in the direction given by $\vec\nabla\zeta^{(\alpha)}_q/\phi_0^{(\alpha)}$ whenever the interspecies interaction $U_{\alpha\beta}$ is negligible. 

If a sphere delimits the boundary of the $\alpha$-fluid there are two alternative configurations. In the case the fluid is contained within the sphere (droplet) or outside it (bubble). Then, singularities at the origin or at infinity are avoided by taking 
$\Lambda^\alpha$ as the adequate solution of Laplace equation in spherical coordinates
\begin{eqnarray}
\Lambda_{\ell m}^\alpha(\vec r) &=&  \mathcal{A}_{\ell m}  Y_{\ell m}(\theta,\varphi) \Big(\frac{r}{\xi^{(\alpha)}}\Big)^\ell\quad\quad \quad{\mathrm{droplet}}  \\
&=&\mathcal{A}_{\ell m}  Y_{\ell m}(\theta,\varphi) \Big(\frac{r}{\xi^{(\alpha)}}\Big)^{-(\ell+1)}\quad {\mathrm{bubble}}, \label{eq:srI}\end{eqnarray}
with $\mathcal{A}_{\ell m}$ a normalization constant to be evaluated according to Eq.~(\ref{eq:nor}) depending on the selection of the $\vec F^\alpha_{\ell m}(\vec r)$ function.
Two reasonable options are identified,

{\bf Ansatz 1}.
\begin{equation} 
\vec F^\alpha_{\ell m}(\vec r) = \mathcal{A}^{(1)}_{\ell m}  Y_{\ell m}(\theta,\varphi)\hat r .\label{eq:a1}
\end{equation}

{\bf Ansatz 2}.
\begin{equation} 
\vec F^\alpha_{\ell m}(\vec r)=  \xi^{(\alpha)}\vec \nabla  \Lambda_{\ell m}^\alpha.
\label{eq:a2}
\end{equation}

We now focus on the particular case  where $\phi_0^{(\alpha)}$ and the external potential $V_{ext}$ do not depend on $\theta$ and $\varphi$.
Then, the corresponding expressions for the $A$, $B_\alpha$ and $C_\alpha$ functionals that determine the excitation energy, Eq.~(\ref{eq:omegaopt}), for a dynamics determined by an extended Gross-Pitaevskii equation are

{\bf Ansatz 1}.
\begin{eqnarray}
A&=&\vert \mathcal{A}^{(1)}_{\ell m} \vert^2\sum_{\alpha=a,b}\xi^{(\alpha)2}\Big[-\frac{1}{2}\int dr \partial_r\rho^{(\alpha)}\partial_r V^\alpha_{ext}
   +\frac{\hbar^2}{2m_\alpha}(\ell(\ell +1)-2\big)\int dr (\partial_r\phi_0^{(\alpha)})^2\Big]\\
B_\alpha&=& -\vert \mathcal{A}^{(1)}_{\ell m} \vert^2\frac{\hbar^2}{2m_\alpha}\ell\int drr\partial_r\rho^{(\alpha)}\Big(\frac{r}{\xi^{(\alpha)}}\Big)^{2\ell} \quad\quad\quad \quad\quad{\mathrm{droplet}}\\ 
&=& \vert \mathcal{A}^{(1)}_{\ell m} \vert^2\frac{\hbar^2}{2m_\alpha}(\ell +1)\int dr r\partial_r\rho^{(\alpha)}\Big(\frac{r}{\xi^{(\alpha)}}\Big)^{-(2\ell+2)} \quad {\mathrm{bubble}}\\
C_\alpha&=& \vert \mathcal{A}^{(1)}_{\ell m} \vert^2\frac{\xi^{(\alpha)}}{2}\int dr r^2\partial_r\rho^{(\alpha)} \Big(\frac{r}{\xi^{(\alpha)}}\Big)^{\ell} \quad\quad \quad\quad\quad\quad{\mathrm{droplet}}\\
&=& \vert \mathcal{A}^{(1)}_{\ell m} \vert^2\frac{\xi^{(\alpha)}}{2}\int dr r^2\partial_r\rho^{(\alpha)} \Big(\frac{r}{\xi^{(\alpha)}}\Big)^{-(\ell+1)} \quad\quad\quad\quad {\mathrm{bubble}}
\end{eqnarray}

{\bf Ansatz 2}. 
\begin{eqnarray}
A&=&\vert \mathcal{A}^{(2)}_{\ell m} \vert^2\sum_{\alpha=a,b}\frac{\xi^{(\alpha)4}}{2}\Big[-\int dr r^2 \partial_r \rho_0^{(\alpha)}\partial_r  V_{ext}(\vec r) \Big(\frac{r}{\xi^{(\alpha)}}\Big)^{2\ell}\nonumber\\
   &+&\frac{\hbar^2}{m_\alpha}\ell^2(\ell -1)(2\ell +1)\int dr r^{-2} (\partial_r\phi_0^{(\alpha)})^2 \Big(\frac{r}{\xi^{(\alpha)}}\Big)^{2\ell}\Big]\quad \quad \quad \quad \quad \quad  {\mathrm{droplet}}\\
&=&\vert \mathcal{A}^{(2)}_{\ell m}\vert^2\sum_{\alpha=a,b}\frac{\xi^{(\alpha)4}}{2}\Big[-\int dr r^2 \partial_r \rho_0^{(\alpha)}\partial_r  V_{ext}(\vec r) \Big(\frac{r}{\xi^{(\alpha)}}\Big)^{-2(\ell+1)}\nonumber\\
   &+&\frac{\hbar^2}{m_\alpha}(\ell +1)^2(\ell +2)(2\ell +1)\int dr r^{-2} (\partial_r\phi_0^{(\alpha)})^2 \Big(\frac{r}{\xi^{(\alpha)}}\Big)^{-2(\ell+1)}\Big]\quad {\mathrm{bubble}}\\
B_\alpha&=& -\vert \mathcal{A}^{(2)}_{\ell m}\vert^2\frac{\hbar^2}{2m_\alpha}\ell\int dr\;r\partial_r \rho_0^{(\alpha)} \Big(\frac{r}{\xi^{(\alpha)}}\Big)^{2\ell}\quad\quad\quad \quad \quad  \quad \quad  \quad \quad \quad \quad \quad  {\mathrm{droplet}}\\
&=& \vert \mathcal{A}^{(2)}_{\ell m}\vert^2\frac{\hbar^2}{2m_\alpha}((\ell+1))\int dr\; r\partial_r \rho_0^{(\alpha)} \Big(\frac{r}{\xi^{(\alpha)}}\Big)^{-2(\ell+1)}\quad \quad \quad \quad \quad \quad  {\mathrm{bubble}}\\
C_\alpha&=& -\frac{m_\alpha\xi^{(\alpha)2}}{\hbar^2} B_\alpha
\end{eqnarray}

For quantum droplets it can be assumed that the ground state order parameters are proportional to each other,
\begin{equation}\label{eq:prop}\phi^{(a)}_0 = \sqrt{\bar{\beta}}\phi^{(b)}_0 = \sqrt{\frac{N^{(a)}}{N^{(b)}}}\phi^{(b)}_0 
\end{equation}
the last equality being a consequence of Eq.~(\ref{eq:normalizacion}).
Since, they can be produced without an external potential, and the ground state depends only on the radial coordinate, the excitation energies become

{\bf Ansatz 1.}
 \begin{equation}\label{eq:energia_ansatz1_esfericas_hetero}
 \hbar^2 \omega^2 \leq
- \ell(\ell-1)(\ell+2)\frac{ \hbar^2}{m_a m_b}
\frac{ N^{(a)}m_b+  N^{(b)} m_a }
     {N^{(a)}m_a  + N^{(b)}m_b}
\frac{ \int dr  (\partial_r \phi^{(a)}_0)^2  
\int dr   \partial_r \rho^{(a)}_0 r^{2\ell+1} 
  }{ \left( \int dr \partial_r \rho^{(a)}_0 r^{\ell+2} \right)^2 } 
  \end{equation}

{\bf Ansatz 2.}
\begin{equation}\label{eq:energia_ansatz2_esfericas_hetero} 
\hbar^2 \omega^2 \leq
   -\ell  (\ell-1)  (2\ell+1)
\frac{ \hbar^2}{m_a m_b}
\frac{ N^{(a)}m_b+  N^{(b)} m_a}
     {N^{(a)}m_a  + N^{(b)}m_b}
      \frac{ \int dr  \,  (\partial_r \phi^{(a)}_0 )^2  r^{2\ell-2}   }
     {  \int dr  \, \partial_r \rho^{(a)}_0 r^{2\ell+1} },
\end{equation}
whenever an equal length scale
\begin{equation}\label{eq:scales} \xi^{(a)} = \xi^{(b)} =\xi,\end{equation}
is chosen for both species. In these equations we notice the presence of the total mass of the droplet
\begin{equation} M_T = N^{(a)}m_a  + N^{(b)}m_b. \end{equation}

In his seminal work \cite{Rayleigh1879}, Lord Rayleigh considered a classical spherical  droplet with constant density of particles $\rho_0$ with mass $m$  and a
radius $R_0$. For the frequency of excitation $\omega_{exc}$, he showed that
\begin{equation}\label{eq:surften}
\omega_{exc}^2 = \ell(\ell -1)(\ell +2)\frac{\sigma}{m\rho_0R_0^3}
\end{equation}  
and identified $\sigma$ as the surface tension.
Lord Rayleigh obtained such an structure for $\omega_{exc}$ analyzing the lowest energy surface excitations of a classical droplet via a variational theorem. The term $$m\rho_0R_0^3 =\frac{3}{4\pi} M_T$$
is directly related to the total mass of the droplet $M_T$.

We observe that the dependence on the multipole parameter $\ell$ both for {\it ansatz} 1 and 2, shows that monopole excitations are not surface excitations, and that the $\ell =1$ contribution is null. This is expected since this term corresponds to a translation of the whole droplet. However for $\ell\ge 2$, the estimation value for $\epsilon =\hbar\omega_{exc}$ of the spherical quantum droplet, and as a consequence of the surface tension, is different for each {\it ansatz}: \begin{itemize}

\item[(i)] For {\it ansatz} 1, it resembles that of Rayleigh and that of the liquid drop model of nuclei in the limit of $N\rightarrow \infty$ nucleons \cite{Bohr1975}. 

\item[(ii)] For {\it ansatz} 2, it is similar to that predicted  for atomic nuclei by Bertsch \cite{berstch1977} and by Casas and Strigari \cite{casas1990} using the random phase approximation and the density-density Green's function formalism; in both cases the second {\it ansatz} was taken for the excitation modes.
\end{itemize}

A variational criteria based on the excitation energy can be applied to select between the two excitation {\it ansatz}.
Following Lord Rayleigh ideas, the corresponding surface tension estimation would be given by the expressions

 {\bf Ansatz 1.}
 \begin{equation}\label{eq:energia_ansatz1_esfericas_hetero}
\sigma^{(1)}_\ell = -\frac{\hbar^2}{M}
\frac{ \int dr  (\partial_r \phi^{(a)}_0)^2  
\int dr   \partial_r \rho^{(a)}_0 r^{2\ell+1} 
  }{ \left( \int dr \partial_r \rho^{(a)}_0 r^{\ell+2} \right)^2 } 
  \end{equation}

{\bf Ansatz 2.}
\begin{equation}\label{eq:energia_ansatz2_esfericas_hetero} 
\sigma^{(2)}_\ell = -\frac{\hbar^2}{M}
      \frac{ \int dr  \,  (\partial_r \phi^{(a)}_0 )^2  r^{2\ell-2}   }
     {  \int dr  \, \partial_r \rho^{(a)}_0 r^{2\ell+1} }
\end{equation}

with $$M = \frac{4\pi }{3}\frac{m_am_b}{N^{(a)}m_b + N^{(b)}m_a}.$$
\section{Extended Gross Pitaevskii equation for binary mixtures of Bose gases.}

Consider a homogeneous weakly repulsive Bose gas composed by hard spheres of diameter ``$\mathrm{a}$" having a density $n$. The  energy correction to the mean field ground state energy introduced by Lee, Huang  and Yang 
\cite{LHY1957} is the first term in a power series expansion in the parameter $(n{\mathrm{a}}^3)^{1/2}$. 
The analogous term for a mixture of two ultracold Bose gases with atom masses $m_a$ and $m_b$, densities $n_a$ and $n_b$, and scattering lengths $\mathrm{a}_{\alpha\beta}$, $\alpha,\beta=a,b$, is
\cite{DErrico2019}:
\begin{equation}\label{funcional_petrov_hetero}
\begin{aligned}
 \epsilon_{LHY} = 
& \frac{1}{4 \pi^2} \left( \frac{m_a}{\hbar^2} \right)^{3/2} (g_{aa} n_a)^{5/2} 
\int^\infty_0 k^2 
\mathcal{F} (k,z,u,x) dk \\
& \mathcal{F} (k,z,u,x) =  \\ 
& \bigg( \frac{1}{2} \left[ k^2 \big( 1 + \frac{x}{z} \big) + 
\frac{k^4}{4}\big( 1+ \frac{1}{z^2} \big) \right] + 
  \bigg[ \frac{1}{4} \big[k^2 + \frac{k^4}{4} - \big( \frac{x}{z} k^2 + \frac{1}{4z^2} \big)   
                     \big]^2    + 
    \frac{ux}{z} k^4 \bigg]^{1/2} \bigg)^{1/2} + \\
& \bigg( \frac{1}{2} \left[ k^2 \big( 1 + \frac{x}{z} \big) + 
\frac{k^4}{4}\big( 1+ \frac{1}{z^2} \big) \right] - 
  \bigg[ \frac{1}{4} \big[k^2 + \frac{k^4}{4} - \big( \frac{x}{z} k^2 + \frac{1}{4z^2} \big)   
                     \big]^2    + 
    \frac{ux}{z} k^4 \bigg]^{1/2} \bigg)^{1/2} \\
& -\frac{1+z}{2z}k^2 - (1+x) + \frac{1}{k^2} \big[1 + x^2z + 4 ux \frac{z}{1+z} \big].
\end{aligned}
\end{equation}
In this equation  the coupling strength factors $g_{\alpha\beta} = 2\pi \hbar^2 \mathrm{a}_{\alpha\beta}/m_{\alpha\beta}$, with $m_{\alpha\beta}=m_\alpha m_\beta/(m_\alpha + m_\beta )$, were introduced as well as the notation,  
$$\frac{m_b}{m_a} = z, \quad  \frac{g^2_{ab}}{g_{aa} g_{bb}} = u,\quad  \frac{g_{bb} n_b}{g_{aa} n_a}=x.
$$
Let us define
\begin{eqnarray}\label{eq:exact}
 f (z,u,x) &=:& \frac{15}{32}  \int^\infty_0 k^2 \mathcal{F} (k,z,u,x) dk  \nonumber\\
&\overset{k = \tan (t)}=& \frac{15}{32}  \int^{\pi/2}_0 \left( \frac{\sin^2(t)}{\cos^4(t)} \right)\mathcal{F} (\tan (t),z,u,x) dt.\label{eq:exact}
\end{eqnarray}
In spite of being a combination of terms that are individually divergent, in general, $f(z,u,x)$ converges~\cite{Ancilotto2018}.
Consider the case $g_{aa}> 0$, $g_{bb} > 0$, and $g_{ab}<0$ and define $\delta g = g_{ab} + \sqrt{g_{aa}g_{bb}}$. Under the weak coupling condition $n\mathrm{a}^3<<1$
it results ~\cite{Petrov2015}  that: (i)the main contribution to the finite integral comes from terms with $k=1/\xi_h =\sqrt{mgn}$; (ii) long-wavelength instabilities expected from a mean field theory approach are cured by quantum fluctuations at shorter wavelengths $k\sim 1/\xi_h \gg \sqrt{m\vert \delta g \vert n}$; (iii) the equillibrum densities are determined by $g_{\alpha \alpha}$, $n_b^{(0)}/n_a^{(0)} = \sqrt{g_{aa}/g_{bb}}$. These conditions define the quantum droplet state. 

\begin{figure}[ht!] %\label{fig:comparacion_F}
            \begin{center}
\includegraphics[width=0.75\textwidth]{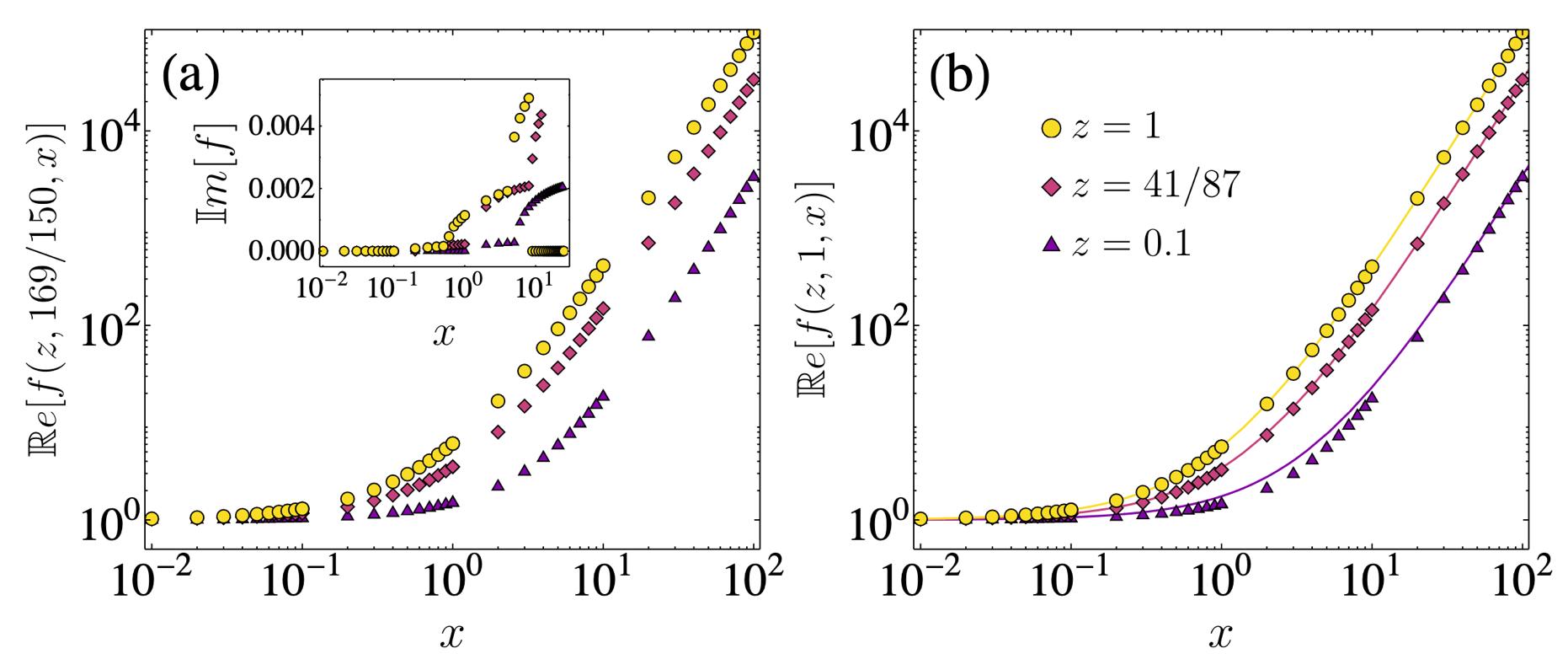}
 \end{center}
   \caption{Illustrative behavior of the complex LHY function $f(z,u,x)$  in the quantum droplet regime $u\sim 1$: (a) $\mathbb{R}e f(z,169/150,x)$ as a function of x and for several $z$ values; inset (a) $\mathbb{I}m f(z,169/150,x)$ as a function of x and for several $z$ values; (b) comparison of  $\mathbb{R}e f(z,u,x)$, using Eq.~(\ref{eq:exact}) (dots),  and its approximate expression Eq.~(\ref{eq:neqmasses}) (solid lines).} \label{fig:comparacion_F}
\end{figure}

For equal masses\cite{Petrov2015}
\begin{equation}
f(1,u,x) = \frac{1}{4\sqrt{2}}\Big[\big(1+x + \sqrt{(1-x)^2 +4ux}\big)^{5/2}+ \big(1+x - \sqrt{(1-x)^2 +4ux}\big)^{5/2}\Big],\label{eq:LHYeqmasses}
\end{equation}
which becomes a complex valued expression for $u < 1$ and  the effective LHY terms yield a non Hermitean EGPE. 
Within the quantum droplets regime $u \sim 1$,  it results that~\cite{DErrico2019}
\begin{equation}\label{eq:neqmasses}
f(z,1,x) \simeq (1+ z^{3/5} x)^{5/2}.
\end{equation}
This approximate expression  for the LHY term in the extended Gross-Pitaevskii equation facilitates the numerical simulations.
However, a direct numerical calculation of $f(z,u,x)$ is also feasible.
In Fig.~\ref{fig:comparacion_F} the results of a numerical calculation of  $f(z,u,x)$ and the result of using the approximate expression Eq.~(\ref{eq:neqmasses}) are illustrated for $u\sim 1$ and several $z$ values, notice the logarithmic scale on (a) and (c) axes.
The value $u=169/150$ in Figs.~\ref{fig:comparacion_F}(a) and (b) approximates that expected for homo--nuclear binary mixtures of $^{39}$K
under feasible experimental conditions~\cite{Cabrera2018,Semeghini2018} ($\mathrm{a}_{aa} = \mathrm{a}_{bb}=48.57 \mathrm{a}_0$, $\mathrm{a}_{ab}= - 51.86 \mathrm{a}_0$), or hetero--nuclear binary mixtures of $^{41}$K  and  $^{87}$Rb, also under feasible experimental conditions~\cite{Fort2021} ($\mathrm{a}_{KK} = 62.0 \mathrm{a}_0$, $\mathrm{a}_{RbRb} = 100.4 \mathrm{a}_0$ and $\mathrm{a}_{KRb}= - 82.0\mathrm{a}_0$). 
 Notice the small relative value of the imaginary part of $f$ with respect to the real one; this fact is frequently used to consider a Hermitean EGPE obtained by just neglecting the imaginary terms derived from $f$. 

The parameters
\begin{equation}
\xi =\hbar\sqrt{\frac{3}{2}\frac{\sqrt{g_{bb}}/m_a + \sqrt{g_{aa}}/m_b}{\vert \delta g\vert \sqrt{g_{aa}}n_a^{(0)}}},\quad
\tau =\frac{3\hbar}{2}
\frac{\sqrt{g_{aa}} + \sqrt{g_{bb}}}{\vert\delta g\vert \sqrt{g_{aa}}n_a^{(0)}}\end{equation}
are identified as the natural  units of length  and time respectively.
In these equations, the saturation density $n_\alpha^{(0)}$ of a quantum droplet in the limit of an infinite number $N$ of atoms for the homo--nuclear case is
\begin{equation}\label{eq:satdenhomo}
n_\alpha^{(0)} = \frac{25\pi}{1024}\frac{ (\mathrm{a}_{ab} + \sqrt{\mathrm{a}_{aa}\mathrm{a}_{bb}})^2}{\mathrm{a}_{aa}\mathrm{a}_{bb}\sqrt{\mathrm{a}_{\alpha\alpha}}(\sqrt{\mathrm{a}_{aa}} + \sqrt{\mathrm{a}_{bb}})^5},
\end{equation}
while for the hetero--nuclear case
\begin{equation}
n_\alpha^{(0)} = \frac{25\pi}{1024} \frac{1}{f^2(m_b/m_a,1,\sqrt{g_{bb}/g_{aa}})}\frac{1}{\mathrm{a}_{\alpha\alpha}^3}\frac{\delta g^2}{g_{aa}g_{bb}}.
\end{equation}
The self-trapping regime requires a minimum number of atoms $N_c^{(\alpha)}$ which can be evaluated by calculating the ground state numerically.
There is a transient regime characterized by a set of  additional critical numbers $N_{ql}^{(\alpha)}$. For $N_{c}^{(\alpha)}<N^{(\alpha)}<N_{ql}^{(\alpha)}$ the atomic density exhibits a surface with a thickness similar to the radius of the droplet. In this regime the concept of
surface excitations is questionable and self-evaporation occurs. For higher $N$ values, the ground state corresponds to that expected for a quantum liquid:  the internal density is almost uniform and the  thickness of its surface is much smaller than the droplet radius. Within the LHY formalism, the values of $N_{c}^{(\alpha)}$ and $N_{ql}^{(\alpha)}$ depend only on the $z,u,x$ variables that determine the $f$ function.

Most of our calculations will be limited to the case $u \sim 1$ using  Eq.~(\ref{eq:neqmasses}), so that the extended Gross-Pitaevskii equation (EGPE) for the binary Bose-Bose mixtures is taken as
\begin{equation}\label{EGPE_hetero--nuclear}
\begin{aligned}
& i \hbar \partial_t \Psi_\alpha = \\  
& \Big( 
- \frac{\hbar^2}{2m_\alpha} \nabla^2 +   g_{\alpha \alpha} |\Psi_\alpha|^2  + g_{\alpha \beta} |\Psi_\beta|^2+ \\
&  \frac{4}{3 \pi^2} \frac{m_\alpha^{3/5} g_{\alpha \alpha}}{ \hbar^3} (m_\alpha^{3/5} g_{\alpha \alpha} |\Psi_\alpha|^2 + m_\beta^{3/5} g_{\beta \beta} |\Psi_\beta|^2 )^{3/2} 
 \Big) \Psi_\alpha.
\end{aligned}
\end{equation}

Within the LHY approximation and a symmetric mixture $N^{(a)}$ = $N^{(b)} = N/2$, $m_a=m_b$ and $\mathrm{a}_{aa}=\mathrm{a}_{bb}$, the equation of state of the droplet reads~\cite{Petrov2015}
\begin{equation}\label{eq:LHYeqM}
\frac{E}{E_0} = -3\Big(\frac{\rho}{\rho_0}\Big) +2\Big(\frac{\rho}{\rho_0}\Big)^{3/2}
\end{equation}
with the equilibrium energy $E_0$ and density $\rho_0$ given by
$$\rho_0 = \frac{25\pi(\mathrm{a}_{aa} + \mathrm{a}_{ab})^2}{1638 \mathrm{a}_{aa}^5}, \quad \frac{E_0}{N} = -\frac{25\pi ^2\hbar^2 \vert \mathrm{a}_{aa} + \mathrm{a}_{ab}\vert^3}{49152m\mathrm{a}_{aa}^5}. $$
Alternative formulations to that of  LHY  corresponds to performing quantum Monte Carlo \cite{Cikojevic2018} or density functional  methods \cite{bottcher2019, Boronat2021} to study dilute Bose-Bose mixtures of atoms with attractive interspecies and repulsive
intraspecies interactions at $T = 0$. Such calculations have been reported for  homo--nuclear mixtures considering diverse finite range interaction potentials. The results are considered to be valid on the bulk, and can be compared to those resulting from LHY in terms of the
equation of state for $u=1$.
Second order density Monte Carlo calculations \cite{bottcher2019} for {\it homo--nuclear} mixtures of atoms in two different hyperfine states give rise to a generalization of Eq.~(\ref{eq:LHYeqM}) with the structure
\begin{eqnarray}
\frac{E}{N} &=& \frac{\vert E_0\vert}{N}\Big[-3\Big(\frac{\rho}{\rho_0}\Big) +\beta\Big(\frac{\rho}{\rho_0}\Big)^\gamma\Big]\label{eq:boro:ab}\\
\beta&=& 1.956\frac{\mathrm{a}_{ab}}{\mathrm{a}_{aa}} + \Big( 0.231 + 0.236\frac{\mathrm{a}_{ab}}{\mathrm{a}_{aa}}\Big) \frac{\mathrm{R}(\mathrm{a}_{ab}, r_{\mathrm{eff}})}{\mathrm{a}_{aa}}\nonumber\\
\gamma&=& 1.83 + 0.32\frac{\mathrm{a}_{ab}}{\mathrm{a}_{aa}} + 0.030\Big( 1 + \frac{\mathrm{a}_{ab}}{\mathrm{a}_{aa}}\Big) \frac{\mathrm{R}(\mathrm{a}_{ab}, r_{\mathrm{eff}})}{\mathrm{a}_{aa}}\label{eq:boro-homo}
\end{eqnarray}
and $\mathrm{R}$ a function which depends on an effective range parameter $r_{\mathrm{eff}}$.  This function can be accessed numerically~\cite{bottcher2019}.  The critical number of atoms $N_{c}^{(\alpha)}$ and $N_{ql}^{(\alpha)}$ will now also depend on $\mathrm{R}$. These calculations have a universal character similar to that inferred from EGPE based on LHY approximation whenever the effective range is included as a parameter.
Using the local density approximation, the following equation for the corresponding order parameter $\Psi$ is found,
\begin{equation}\label{eq:boro}
\begin{aligned}
& i \hbar \partial_t \Psi (\vec{x})=   
 \Big( 
- \frac{\hbar^2}{2m} \nabla^2  + \frac{25 \pi^2 \hbar^2  \vert\mathrm{a}_{aa} + \mathrm{a}_{bb} \vert^3}{49152m\mathrm{a}_{aa}^5} \times  \\
&\Big[ -6 \frac{N^2 |\Psi|^4}{\rho_0}
+\beta (\gamma+1) \left( \frac{  N |\Psi|^2 }{\rho_0} \right)^\gamma
\Big] \Big) \Psi(\vec{x}).
\end{aligned} 
\end{equation}
For hetero-nuclear quantum droplets, a generatization of Eq.~(\ref{eq:boro:ab}) has been reported in Ref.~\cite{Boronat2021} for Rb-K mixtures and particular values of the $\mathrm{a}_{\alpha\beta}$ coupling parameters.

\subsection{Droplets formed by binary mixtures of  homo--nuclear atoms.}

The ground state of the EGPE in the droplet regime 
is expected to exhibit~\cite{Petrov2015},
\begin{itemize}
\item[(i)] a saturation density given by  $n_\alpha^{(0)}$, Eq.~(\ref{eq:satdenhomo}), in the limit of $N^{(\alpha)}\rightarrow \infty$; then,  $n_\alpha^{(0)}$ becomes a natural unit to measure the droplet density of  $\alpha$-species;

\item[(ii)] a spherical shape with radius $R_0\approx ( 3 N/4\pi n_\alpha^{(0)}\xi^3)^{1/3}\xi$, 
for finite $N^{(a)} = N^{(b)} = N/2$;

\item[(iii)] a surface thickness  of order $\xi$. 

\end{itemize}

The well-known shape of a  Boltzmann function 
\begin{equation}
\rho_B(R;N) = \frac{A_1}{1 + \mathrm{exp}((R-R_0)N/dR)}\label{eq:Boltzmann}
\end{equation}
makes of it a feasible mathematical model of the spherical density profile of a droplet for a given number of particles. Its parameters $R_0$ and $dR$ are then 
interpreted as the radius of the droplet and its surface thickness respectively.
The saturation density,
\begin{equation}
\rho_B(0;N) = \frac{A_1}{1 + \mathrm{exp}(-R_0 N/dR)}
\end{equation}
is required to fix the $A_1$ parameter. The function $\rho_B(R;N)$ will be used
to get approximate fits to the density profiles obtained numericaly from EGPE and from the effective range  model.
\begin{figure}[h!] 
\includegraphics[width=0.75\textwidth]{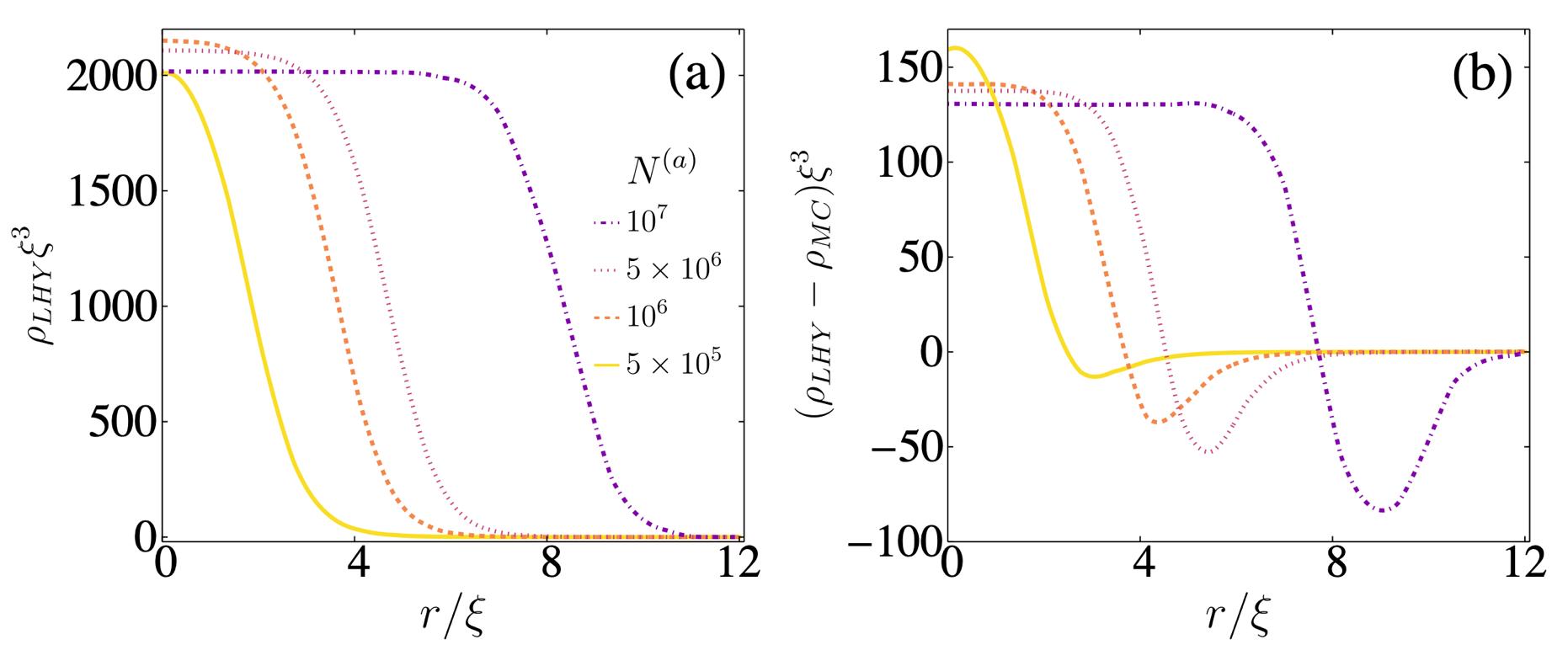}
\caption{ \footnotesize{(a)Ground state densities $\rho_{\mathrm{LHY}}$ as a function of the radial distance for a binary mixture of homo--nuclear atoms evaluated using the functional Eq.~(\ref{funcional_petrov_hetero}); (b) corrections to the LHY mean field predicted by Monte Carlo inspired calculations, Eq.~(\ref{eq:boro}) and evaluated by substracting the corresponding ground state density $\rho_{\mathrm{MC}}$ from the LHY prediction $\rho_{\mathrm{LHY}}$.
In this illustrative example, $a_{aa} = 48.57a_0 , a_{ab} = −51.86a_0 , \gamma = 1.48878$, and $\beta = 2.08519$. The natural unit of length takes the value
$\xi = 1.455 \mu m$.} \label{fig:densities}}
\end{figure}

We focus first on the symmetric, $N^{(a)}=N^{(b)}=N/2$, homo--nuclear $m_a=m_b$ case, and compare the results derived from the  EGPE resulting from LHY and from the effective range model derived from the Monte Carlo density    simulations. Taking into account the experimental realizations we consider
$^{39}$K atoms and scattering lengths compatible with the Feshbach resonances of such atoms. In the illustrative examples reported in this Section, those scattering lengths
are $\mathrm{a}_{aa} = 48.57 \mathrm{a}_0$, $\mathrm{a}_{ab}= - 51.86 \mathrm{a}_0$. 

The  density profiles resulting from EGPE equations are illustrated in Fig.~\ref{fig:densities}a.  The droplet radius  obtained from fitting a Boltzmann distribution satisfy the equation
\begin{eqnarray}
R_0(N^{(a)}) &=& ((-.54\pm 0.01) +(0.0518\pm 0.0002)\times N^{(a)1/3})\xi,\quad  N^{(a)}> N_c^{(a)},\\
       & = &( (-.54\pm 0.01) +(1.051\pm 0.004)\times \big(3N /4\pi n_0^{(a)}\xi^3\big)^{1/3})\xi.
\end{eqnarray}
Our numerical simulations reveal that the number of atoms required to exist a self bound state is $N^{(a)}_c\approx 18.65 n_0^{(a)}\xi^3 \approx 37210 $ in accordance to the results reported in Ref.~\cite{Petrov2015}. 
The saturation density is given by
\begin{equation}
\rho_B(0;N^{(a)}) = ((0.95\pm 0.17) + (0.24\pm 0.38)\, \mathrm{exp}(-(N^{(a)}/2)^{1/3})n_0^{(a)},
\quad  N^{(a)}\ge 3N^{(a)}_c,
\end{equation}
 and the thickness of the surface results almost independent on $N^{(a)}$ for $3N^{(a)}_c <N^{(a)}\le 10^7$ and given by
\begin{equation}
dR = (0.53\pm 0.01)\xi. 
\end{equation}
 The resulting density profiles fits the Boltzmann density  with a reliability above 0.998  for droplets with $N > N_c$ and even higher reliabities are found for $N>3N_c$. The predictions on the properties of the ground state that were mentioned in Ref.~\cite{Petrov2015} for the saturation density, and the general behaviour of $R_0$ and $dR$ within EGPE, have been numerically tested with the results that we have illustrated. 

A similar analysis for the ground state infered from effective range simulations based on Eq.~(\ref{eq:boro:ab}) yields 
\begin{eqnarray}
R_0(N) &=& ((-.54\pm 0.03) +(0.0527\pm 0.0003)\times N^{(a)1/3})\xi,\quad  N^{(\alpha)}> N_c^{(\alpha)},\\
 & = &( (-.54\pm 0.03) +(1.069\pm0.006)\times \big(3N^{(\alpha)} /4\pi n^{(0)}\xi^3\big)^{1/3})\xi,
\end{eqnarray}
in this case $N_c^{(a)}\approx 40000$.
It results that the ground state exhibits a saturation density that satisfies the equation
\begin{equation}
\rho_B(0;N^{(\alpha)}) = (0.907\pm 0.005) + (0.34\pm 0.26)\, \mathrm{exp}(-(N^{(\alpha)}/2)^{1/4}) n_0^{(a)},
\quad  N^{(\alpha)}\ge 3\cdot N_c^{(\alpha)},
\end{equation}
and a mean thickness of the surface 
\begin{equation}
dR = (0.56\pm 0.02)\xi ,\quad N^{(\alpha)}>N_c^{(\alpha)}.
\end{equation}
In this case, the thickness has a slightly non monotonic dependence on $N^{(\alpha)}$, and it is in the interval (0.55,0.57)$\xi$ for $10^5\le N^{(\alpha)}\le 10^7$. The saturation density is slightly lower for the effective range simulations with a corresponding increase in both the value of $R_0$ and $dR$. Notice also a slightly different dependence on $N^{(\alpha)}$ for the saturation density.

\subsubsection{Excitation energies.}
\begin{figure}[h!]
 \begin{center}
\includegraphics[width=0.75\textwidth]{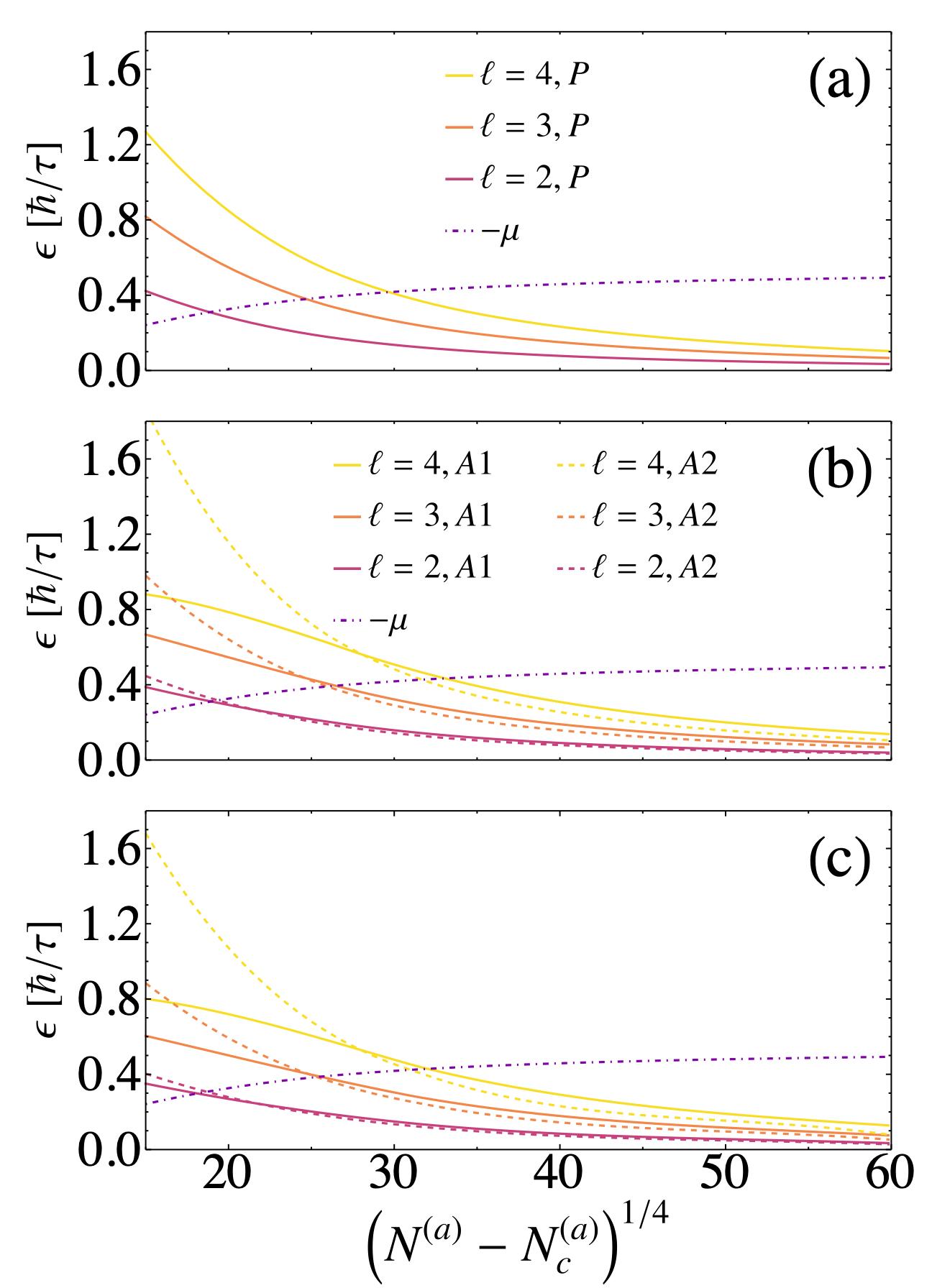}
\end{center}
\caption{ Approximate excitation energies $\epsilon_\ell$ as a function of the number of atoms $N$ for an homo--nuclear mixture of $^{39}$K atoms.  The negative value of the chemical potential $\mu$ is also shown (dot-dash). The excitation energies are evaluated 
according to different models: (a-b) densities evaluated within the EGPE; (c) densities evaluated using the effective interaction Eq.~(\ref{eq:boro:ab}). In (a) the phenomenological expression Eq.~(\ref{eq:pheno}) is used; in (b-c)the results from  {\it ansatz} 1, A1, (continuos) and {\it ansatz} 2, A2, (dashed) are shown.  The natural  unit $\hbar/\tau$ of energy is used with $\tau =1.308$ms. In all graphs $\ell=2$ corresponds to green lines, $\ell=3$ to red lines and $\ell=4$ to blue lines. $N_c^{(a)}\sim 37210$ for the EGPE and $N_c^{(a)}\sim 45000$ for the Monte Carlo inspired calculations.
}\label{fig:exci:homo}
\end{figure}

The ground state wave functions both for the EGPE and the MC inspired  effective range equations can be used to calculate the  energies associated to surface excitations.  Within a description where the effective dynamical equations consider just contact interactions, Eq.~(\ref{eq:energia_ansatz1_esfericas_hetero}) and Eq.~(\ref{eq:energia_ansatz2_esfericas_hetero}) taking $m_1 = m_2$,
give an estimation to those energies according to each {\it ansatz}. 
The numerical results are illustrated in Fig.~\ref{fig:exci:homo}, where they are compared to the analytical expression
\begin{equation}\label{eq:pheno}
\tilde{\epsilon}_\ell^2 = \frac{4\pi(1 + \sqrt{3})}{35}\ell(\ell -1)(\ell +2)\frac{n_i^{(0)}\xi^3}{N}.
\end{equation}
According to Ref.~\cite{Barranco2016} $\tilde{\epsilon}_\ell$ should provide a precise estimation of the excitation energy  for $N\gg 1$. The expression of $\tilde{\epsilon}_\ell$ is based in the random phase approximation without a variational support.
Taking into account the variational nature of  $ansatz$ 1 and 2, the latter gives a better estimation of the energy for small values of $N$. However, for a given $\ell$ value, there is a critical value $N_\ell^c$ for which {\it ansatz} 1 gives a lower excitation energy if $N >N_\ell^c$. The third phenomenological estimation of the excitation energies obtained from Eq.~(\ref{eq:pheno}) cannot be considered closer to the exact one eventhough it may be lower than the result of {\it ansatz} 1 or 2, both because it is not supported by a variational theorem and because it is  expected to be valid just for $N>>1$; note that  the {\it ansatz} 1 results coincide with Eq.~(\ref{eq:pheno}) in this limit.

For low $N$ calculations where {\it ansatz} 2 gives lower energies, droplets with a radius $R_0$ similar to its width dR result, there the quantum incompressible regime has not been reached. This observation is  consistent with the $\ell$ dependence of the energy involving the factor $\ell(\ell-1)(\ell +2)$ that was  predicted  by Lord Rayleigh for classical liquids, and which is also predicted by {\it ansatz} 1 and the phenomenological result Eq.~(\ref{eq:pheno}), but not by {\it ansatz} 2.

As illustrated in Figure~3, the similarities between the results for the excitation energies obtained for LHY calculations and the effective equation resulting from
density Monte Carlo calculations are remarkable. 

The numerical evaluation of the surface excitations allows the identification of the surface tension according to Eq.~(\ref{eq:surften}). For excitations departing from the real ground state function that have a Boltzmann function shape, Eq.~(\ref{eq:Boltzmann}), the surface tension  Eqs.~(\ref{eq:energia_ansatz1_esfericas_hetero}-\ref{eq:energia_ansatz2_esfericas_hetero}) can be written in terms of the Fermi-Dirac integrals 
\begin{equation}
F_s (z)= \frac{1}{\Gamma(s+1)}\int _0^\infty \frac{dx x^s}{1+e^{x-z}}.
\end{equation}
So that, for $\ell> 1$
\begin{eqnarray}
\sigma_\ell^{(1)} &=& \frac{\hbar^2}{8M dR^4}\frac{(2\ell +1)!}{((\ell +2)!)^2}\Big[1-\frac{1}{(1+ e^{R_0/dR})^2}\Big]
                      \frac{F_{2\ell}(R_0/dR)}{(F_{\ell +1}(R_0/dR))^2}, \\
\sigma_\ell^{(2)} &=& \frac{\hbar^2}{16M dR^4}\frac{1}{\ell(4\ell^2 -1)}
                      \frac{F_{2\ell-3}(R_0/dR)+F_{2\ell -4}(R_0/dR)}{F_{2\ell}(R_0/dR)},
\end{eqnarray}
expressions that correspond to  {\it ansatz} 1 and 2 respectively.

\subsection{Droplets formed by  binary mixtures of hetero--nuclear atoms.\label{sec.groundhetero}}

Experimental results have already been reported for mixtures of $^{41}$K and $^{87}$Rb \cite{Fort2021}, for scattering lengths $\mathrm{a}_{KK} =62.0\mathrm{a}_0$, $\mathrm{a}_{RbRb} =100.4\mathrm{a}_0$, respectively. In those experiments,  the interspecies scattering length  was explored in the interval $\mathrm{a}_{KRb} \epsilon (-95,-72)\mathrm{a}_0$, and the observed binary mixtures satisfy $N^{(Rb)}/N^{(K)} \sim 1.15$. In the numerical illustrative examples along this manuscript, we take $\mathrm{a}_{KRb}=-82.0\mathrm{a}_0$. Under these conditions the natural scales for length are $\xi^{(K)}= \xi^{(Rb)}= 1.147\mu$m = $\xi$, and for time $\tau^{(K)}= \tau^{(Rb)}= 1.191\cdot 10^{-3}$s = $\tau$.
The ideal saturation densities are $n_0^{(K)} =3.745\cdot 10^{20} \mathrm{m}^{-3}$ and  $n_0^{(Rb)} =4.287\cdot 10^{20} \mathrm{m}^{-3}$,
in fact  $n_0^{(Rb)}/n_0^{(K)} = \sqrt{g_{KK}/g_{RbRb}}$. 

The ground state density was obtained by solving the EGPE within the LHY approach,Eq.~(\ref{EGPE_hetero--nuclear}).  It must be mentioned that the difference between the predictions of the LHY approach and the density Monte Carlo formalisms are expected to increase as $\vert \mathrm{a}_{KRb}\vert$ increases \cite{Boronat2021}
taking the smallest values for $-85\mathrm{a}_0< \mathrm{a}_{KRb}< -77\mathrm{a}_0 $ whenever the  coupling parameters $a_{RbRb}$ and $a_{KK}$ are similar to those mentioned in the former paragraph.
 As a first step in our calculations,  the critical number of particles required to achieve self--trapping was found to be $N_c^{(K)}\sim 11000$ which is close but slightly greater to the expectation $N^{(K)}_{homo} =18.65\times n_0^{(K)}\xi^{(a)3}=10535 $ infrerred from the theoretical description of  homo--nuclear mixtures. The structure of the ground state densities was then studied using
the Boltzmann fitting curve Eq.~(\ref{eq:Boltzmann}). In general the reliability of the fitting was above 0.998 increasing along the number of atoms. The radius of the quantum droplets as a function of the number of atoms $N^{(\alpha)}$ satisfy the equations
\begin{eqnarray}
R^{(K)}_0(N^{(Rb)})& = &(( -0.471\pm 0.018) +(0.07306\pm 0.00018)\times N^{(Rb)1/3})\xi,\quad  N^{(Rb)}\ge N_c^{(Rb)},\\
       & = &( (-.471\pm 0.018) +(0.973\pm 0.002)\times \big(3N^{(K)} /4\pi n_0^{(K)}\xi^3\big)^{1/3})\xi,
\end{eqnarray}
for $^{41}$K atoms, and
\begin{eqnarray}
R^{(Rb)}_0(N^{(Rb)})& = &((-0.480\pm 0.018) +(0.07309\pm 0.00018)\times N^{(Rb)1/3})\xi,\quad  N^{(Rb)}> N_c^{(Rb)},\\
       & = &( (-0.480\pm 0.018) +(0.975\pm 0.002)\times \big(3N^{(Rb)} /4\pi n_0^{(K)}\xi^3\big)^{1/3})\xi,
\end{eqnarray}
for $^{87}$Rb atoms. 
The thickness $dR$ results almost independent on $N^{(\alpha)}$ for $3N^{(\alpha)}_c <N^{(\alpha)}\le 10^7$, and it is given by
\begin{equation}
dR = (0.52\pm 0.01)\xi. 
\end{equation}
For $N^{(\alpha)}_c <N^{(\alpha)}< 3N^{(\alpha)}_c$, $dR$  decreases as $N^{(\alpha)}$ increases being always in the interval $ dR$  $\epsilon$  $[0.51\xi,0.61\xi]$.

\begin{figure}[h!]
 \begin{center}
\includegraphics[width=0.75\textwidth]{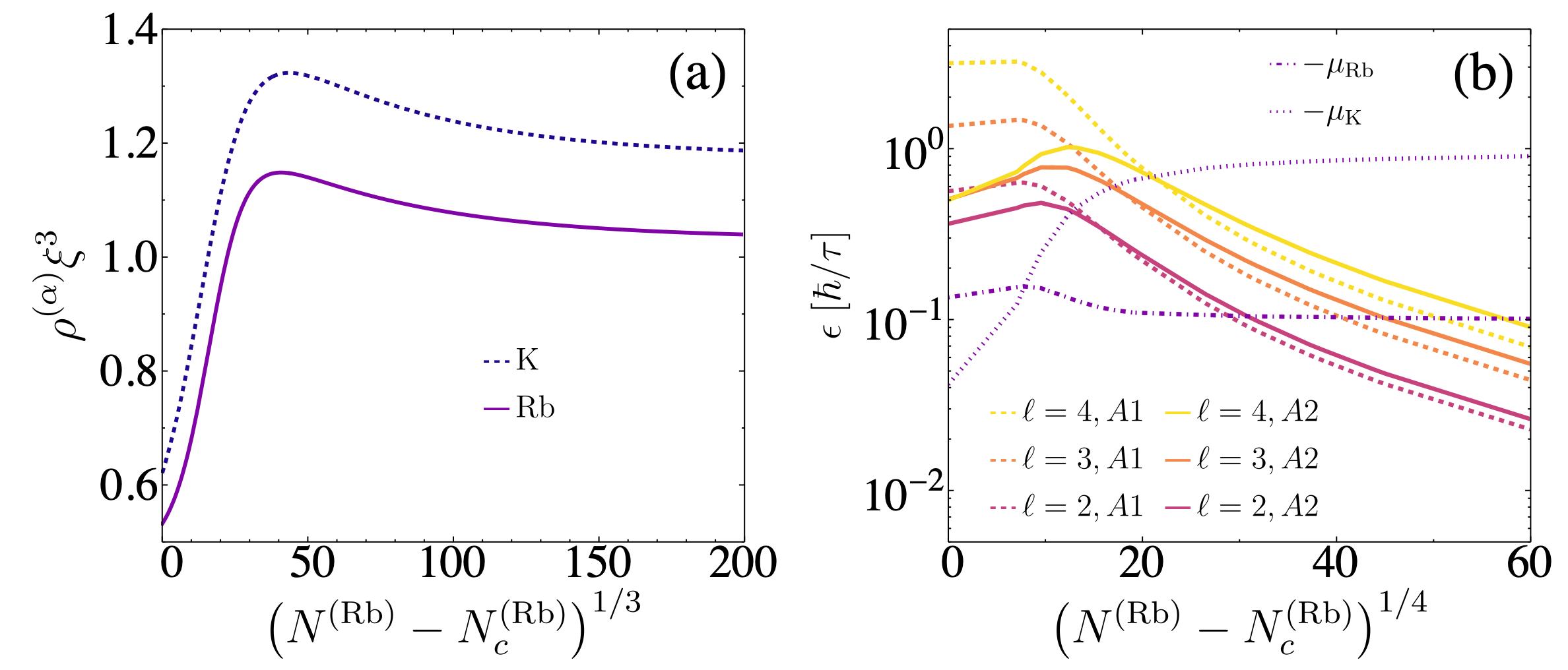}
\end{center}
\caption{(a) Saturation densities of $^{41}$K (dashed-blue) and  $^{87}$Rb (continuos-purple) as a function of the number of $^{87}$Rb atoms $N^{(Rb)}$. (b) Approximate excitation energies $\epsilon_\ell$ as a function of the number of atoms $N^{(Rb)}$ and the angular momentum number $\ell$. They are  evaluated 
using  the ground state densities obtained from the EGPE; ansatz~1 (dashed) and ansatz~2 (continuos).  The negative value of the chemical potential $-\mu_K$ (purple, dot-dash) and $-\mu_{Rb}$ (purple, dotted)  are also shown. The natural energy unit $\hbar/\tau$ is used with $\tau =1.191$ ms. }\label{fig:satdenhetero}
\end{figure}

The saturation densities exhibit a non monotonic behavior as a function of the number of atoms. This is illustrated in Fig.~\ref{fig:satdenhetero}a. The numerical results also show that, for a given droplet both species share the same radius and the saturation densities satisfy $\rho_b(0,\infty)/\rho_a(0,\infty) = \sqrt{g_{aa}/g_{bb}}$. Since the order parameters $\Psi^{(\alpha)}$
are real, these results are  numerically congruent with the proportionality between those densities, Eq.~(\ref{eq:prop}).
They can be approximately described by the equation,
\begin{equation}
\rho_\alpha(0;x^{(\alpha)})= A^{(\alpha)}_2+\frac{(A^{(\alpha)}_1-A^{(\alpha)}_2) + \mathrm{e}^\frac{x^{(\alpha)}-x^{(\alpha)}_0}{dx^{(\alpha)}_0}}{1+\mathrm{e}^\frac{x^{(\alpha)}-x^{(\alpha)}_1}{dx^{(\alpha)}_1}}
\end{equation}
with $x^{(\alpha)} = (N^{(\alpha)}- N_c^{(\alpha)})^{1/3}$, the fitting parameters are given in Table~ \ref{table:satden} for the system here exemplified. They quantify the characteristic
steep exponential growth of $\rho_\alpha(0;x)$ for small $x$ values followed by soft decrease after the maximum of the saturation density is achieved. The latter
occurs at $(N^{(Rb)}-N_c^{(Rb)})^{1/3} \approx 42.5$, equivalent to $~$89435$\approx 7 N_c^{(Rb)}$ Rb atoms.
For a smaller  $N^{(Rb)}$, the droplet radii $R_0\sim dR$, so that the interpretation of the latter as an effective width of the droplet surface is not evident.
That is, for the system under consideration the region where self confinement is achieved and the fluid is compressible corresponds to
$N_c^{(\alpha)}\le N^{(\alpha)}\le 7N_c^{(\alpha)}$. For higher $N^{(\alpha)}$ the fluid is approximately incompressible.

\begin{table}[ht]
\caption{Fitting coefficients for the saturation density.}
\centering 
\begin{tabular}{|c |c| c| c| c| c| c|}
\hline
\hline                       
Atom       & $A_1^{(\alpha)}$& $A_2^{(\alpha)}$ & $dx_0^{(\alpha)}$ &  $dx_1^{(\alpha)}$ &  $x_0^{(\alpha)}$ &  $x_1^{(\alpha)}$\\
\hline 
$^{41}$ K &  1.032 $\pm$ 0.002 & 0.475 $\pm$ 0.004 & -55.9 $\pm$ 0.5 & -6.36 $\pm$ 0.11 & -72.7 $\pm$ 1.1 & 16.59 $\pm$ 0.14\\
\hline
$^{87}$ Rb & 1.177 $\pm$ 0.003 & 0.450 $\pm$ 0.005 & -54.6$\pm$ 0.6 & -8.39 $\pm$ 0.24 & -52.3 $\pm$ 1.2 & 14.3 $\pm$ 0.6\\
\hline
\end{tabular}
\label{table:satden}

\end{table} 
The behaviour of the two chemical potentials is different for the two species as expected from their differences in mass and  coulping interactions.
In Fig.~\ref{fig:satdenhetero}b, the values of the negative of the chemical potentials $-\mu_K$ and $-\mu_{Rb}$ of the ground states obtained from EGPE are shown. Notice that $\vert \mu_K\vert$ always increases as $N^{(K)}=\sqrt{g_{RbRb}/g_{KK}} N^{(Rb)}$ increases, while for $^{87}$Rb atoms $\vert \mu_{Rb}\vert$ at $N^{(Rb)}=N_c^{(Rb)}$ is higher than its asymptotic value$~ 0.1\hbar/\tau$ for $N^{(Rb)}\rightarrow \infty$.
The crossing point of those chemical potentials corresponds to $N^{(Rb)}-N_c^{(Rb)} \approx 8^4$, that is $N^{(Rb)}\approx 16326\approx 1.3\cdot N_c^{(Rb)} $ which is smaller than the value at which the maximum of the saturation density is achieved. That is, for $N_c^{(\alpha)}\le N^{(\alpha)}\le 1.3N_c^{(\alpha)}$
it is energetically favorable to release a single $^{41}$K atom from the fluid, than a  $^{87}$Rb atom; the reverse condition applies for higher $N^{(\alpha)}$ values.

\subsubsection{Excitation energies.}
In Fig.~\ref{fig:satdenhetero}b, the excitation energies evaluated using {\it anstaz} 1 and 2 are shown. In a similar way than the homo--nuclear case, {\it ansatz} 2 gives a better estimation of the energy for the compressible regime at small values of the number of atoms, $N^{(Rb)}< 70227$ for quadrupole excitations and $N^{(Rb)}< 10^7$ for octupole excitations.

 {\it Ansatz} 1 must be taken as a better variational estimation for high $N^{(a)}$ values. Whenever an excitation energy is higher than a chemical potential, self-evaporation is expected to occur. The differences in the value of the chemical potential indicates that the self-evaporation rates of each species differ and depend on the number of atoms in the droplet.

\subsubsection{Ideal evolution of the excitation modes.\label{Sec:ideal}}
Within the Bogolubov scheme, the  excitation modes of the quantum droplets are determined by the normalized and complete basis set $\{u_q^{(a)},v_q^{(a)},u_q^{(b)},u_q^{(b)}\}$.
An ideal collective excitation  of the wave function $\psi^{(\alpha)}_{exc} $ shows a harmonic time dependence 
\begin{equation}
\psi^{(\alpha)}_{exc} (\vec r,t)= e^{-i\mu_\alpha t/\hbar}\Big(\phi^{(\alpha)}_0(\vec r) +
\sqrt{N^{(\alpha)}}(u_q^{(\alpha)}(\vec r) e^{-i\omega_q t} + v_q^{(\alpha)*}(\vec r) e^{i\omega_q t}) 
\Big)\label{eq:single-exc}.
\end{equation}
for a single value of the parameters $q$ that characterize it. 
In this Section we illustate the atomic densities for hetero--nuclear mixtures resulting directly from this equation and applied to the ground state wave function $\phi^{(\alpha)}_0$ of the EGPE being $u_q^{(a)}$ $v_q^{(a)}$ the functions determined by either {\it ansatz} 1 and 2. The results facilitate the description of the consequences of the collisions between quantum droplets which are reported in Section  V.

We focus on ideal quadrupole excitations and consider two cases of particular relevance. The first corresponds to the  minimum number of atoms $N_c^{(\alpha)}$ for which self--trapping is expected, and
the second to the number of atoms at which the excitation energy equals the negative of the chemical potential. Graphs are obtained from isodensity surfaces chosen to approximately delimit the droplets.
Notice that for $N\sim N_c^{(\alpha)}$ the deformation of the droplet with respect to the ground state as a function of time is less pronounced for {\it ansatz} 2 than for {\it ansatz} 1. This is congruent with the compressible character of the quantum fluid for this number of atoms, and  gives a qualitative understanding to the lower excitation energy of this {\it ansatz} in that regime. For $N^{(\alpha)}\sim 3.5 N_c^{(\alpha)}$ the comparative evolution of the droplets is similar at several times. Apparently the droplets seem to divide in several sub-droplets during particular  time intervals. Notice however this interpretation is not precise if it is just based in the isodensity surfaces. A more detailed analysis beyond the illustrative isosurfaces shows that there is always a dilute gas between the apparent fragments.

\begin{figure}
 \begin{center}
\includegraphics[width=0.85\textwidth]{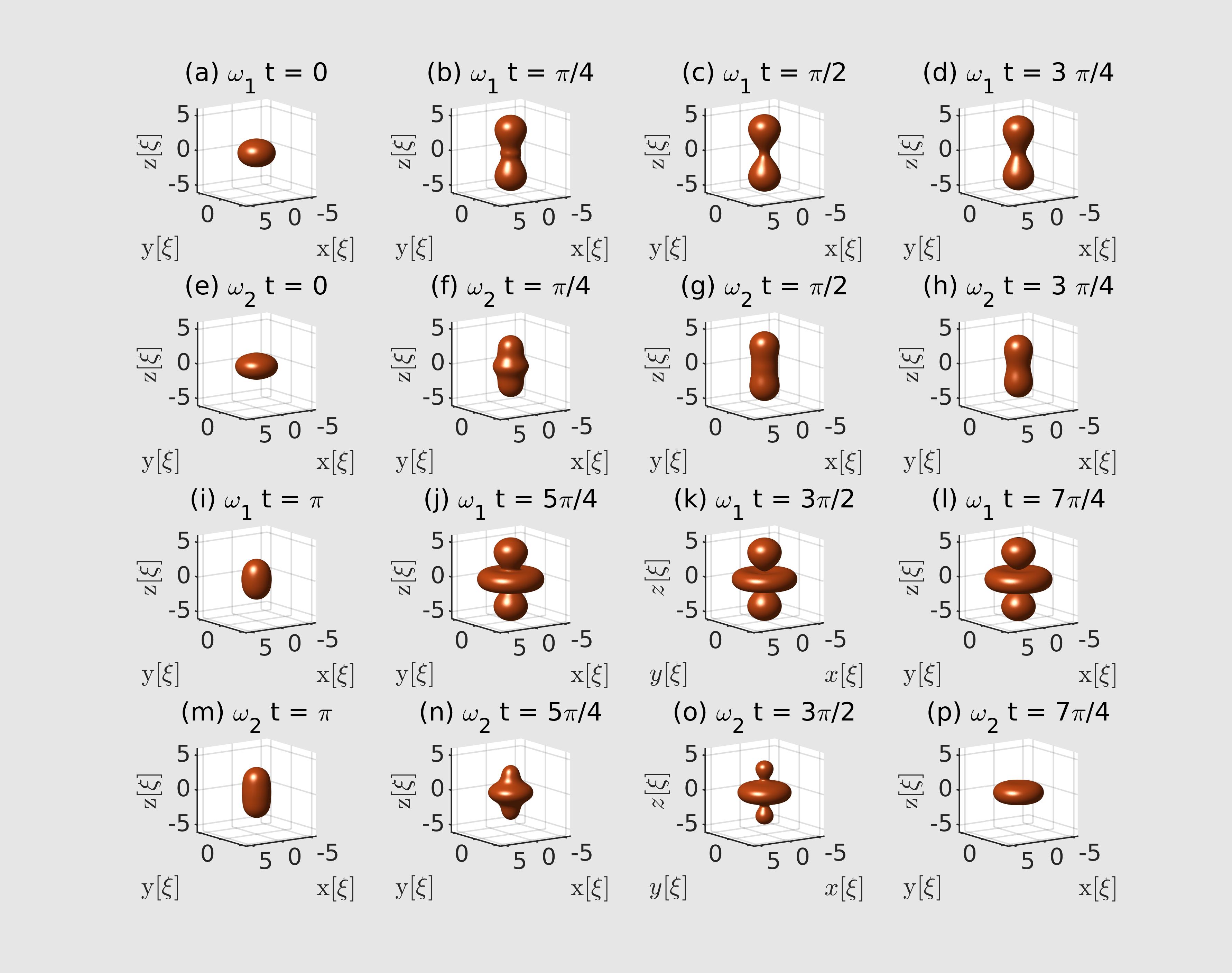}
\end{center}
\caption{Comparative illustrations of the ideal evolution of the density of a quantum droplet formed by $N^{(Rb)}$ = 12699  and $N^{(K)}$ = 11016
atoms considering the excited quadrupole state with $m_\ell =0$  generated from the ground state droplet.
{\it Ansatz} 1  corresponds to rows (a-d) and (i-l)) and {\it ansatz} 2 to rows (e-h) and (m-p). The 3D graphs where generated by plotting isodensity surfaces at fifteen percent of the maximum density at each time.}\label{fig:excprofiles-12699}
\end{figure}
\newpage

\begin{figure}
 \begin{center}
\includegraphics[width=0.85\textwidth]{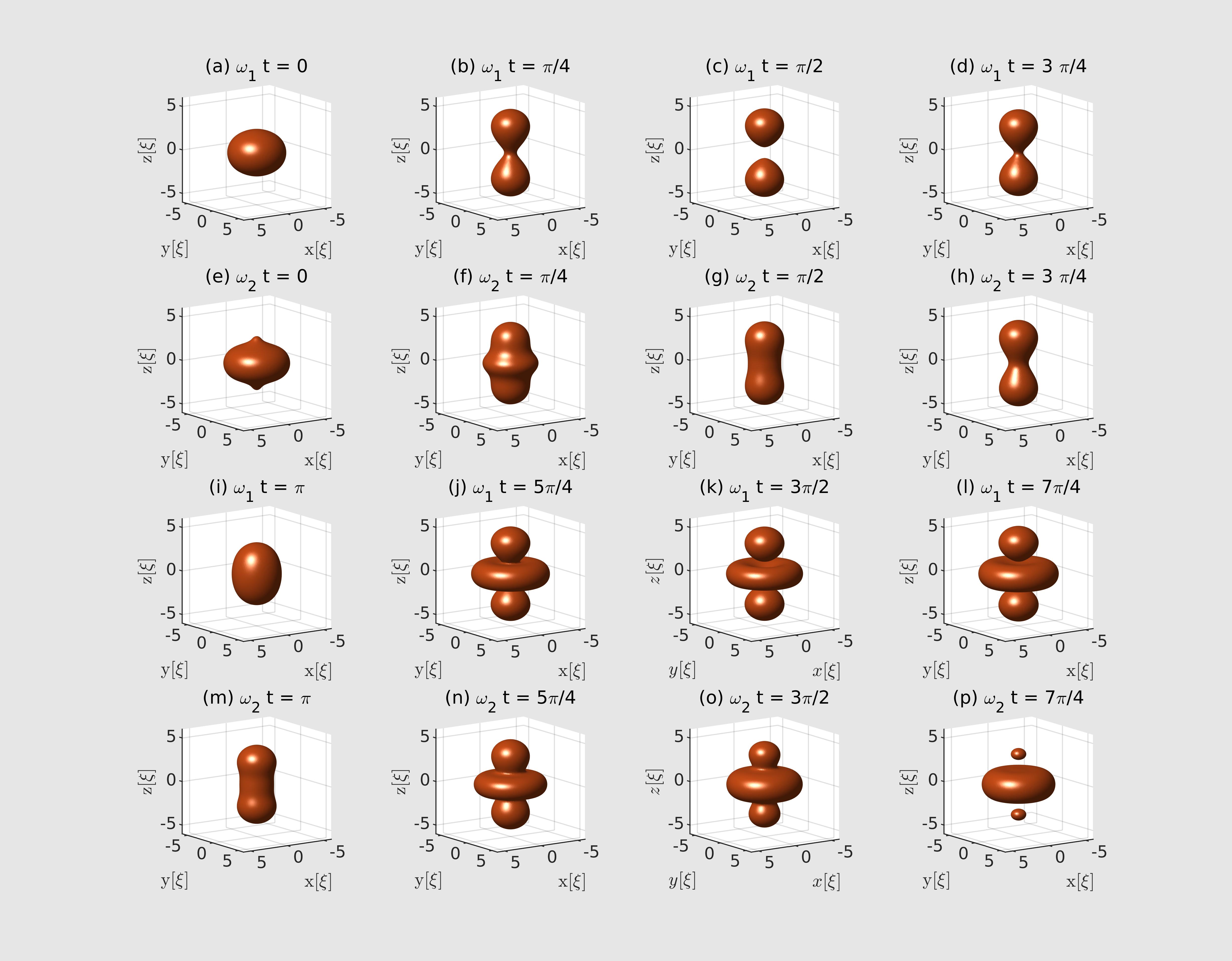}
\end{center}
\caption{Comparative illustrations of the ideal evolution of the density of a quantum droplet formed by $N^{(Rb)}$ = 43636 and $N^{(K)}=37944$ atoms corresponding to the excited quadrupole state with $m_\ell =0$  generated from the ground state droplet.
{\it Ansatz} 1  corresponds to rows (a-d) and (i-l) and {\it ansatz} 2 to rows (e-h) and (m-p). The 3D graphs where generated by plotting isodensity surfaces at fifteen percent of the maximum density at each time.}\label{fig:excprofiles-43636}
\end{figure}

\section{Atoms losses.} 

Both for homo--nuclear and hetero--nuclear mixtures, if the number of atoms is low enough the excitation of normal modes for the droplets is energetically more expensive than the loss of atoms in the droplets.
This self-evaporation process may prevent the observation  of the ideal modes discused in previous Sections. Notice that for hetero--nuclear mixtures, the rates of evaporation for each species  differ. This could lead to a second mechanism that threatens the stability of the droplet since the quotient $N^{(a)}/N^{(b)}$ could be altered as well as the  high overlap of the two atomic wave functions
that is required for self-trapping. Finally, three--body scattering that has not been included in the EGPE can modify the internal state of the atoms, giving rise to a third instability mechanism.

In the following Subsections, these mechanisms are studied. We are particularly interested in the evaluation of the expected losing rates for the mixtures we have studied. Notice that in all experiments reported in the literature, these rates have conditioned the possibility of observing both the ground state droplets for long times and large number of atoms, and the excitations induced by, for instance, droplets collisions.  

\subsection{Self evaporation}

The evaluation of self evaporation is directly connected with the excitation of a quantum droplet. For its evaluation, we calculated numerically the evolution of an excited droplet using the time dependent EGPE. The initial state of the atomic cloud is again taken as given by Bogolubov expression,
Eq.~(\ref{eq:single-exc}) at $t=0$. As difference to the Section~\ref{Sec:ideal} study, the evolution of the excitation is here obtained by solving numerically the EGPE. The evaporation rates are evaluated numerically
in terms of the atoms that leave the droplet as a function of time; the latter are identified as those located outside the minimum radius of the sphere that includes all the atoms in the ideal evolution of the approximate excitation mode given by Bogolubov expression.

\begin{figure}[h!]
 \begin{center}
\includegraphics[width=0.75\textwidth]{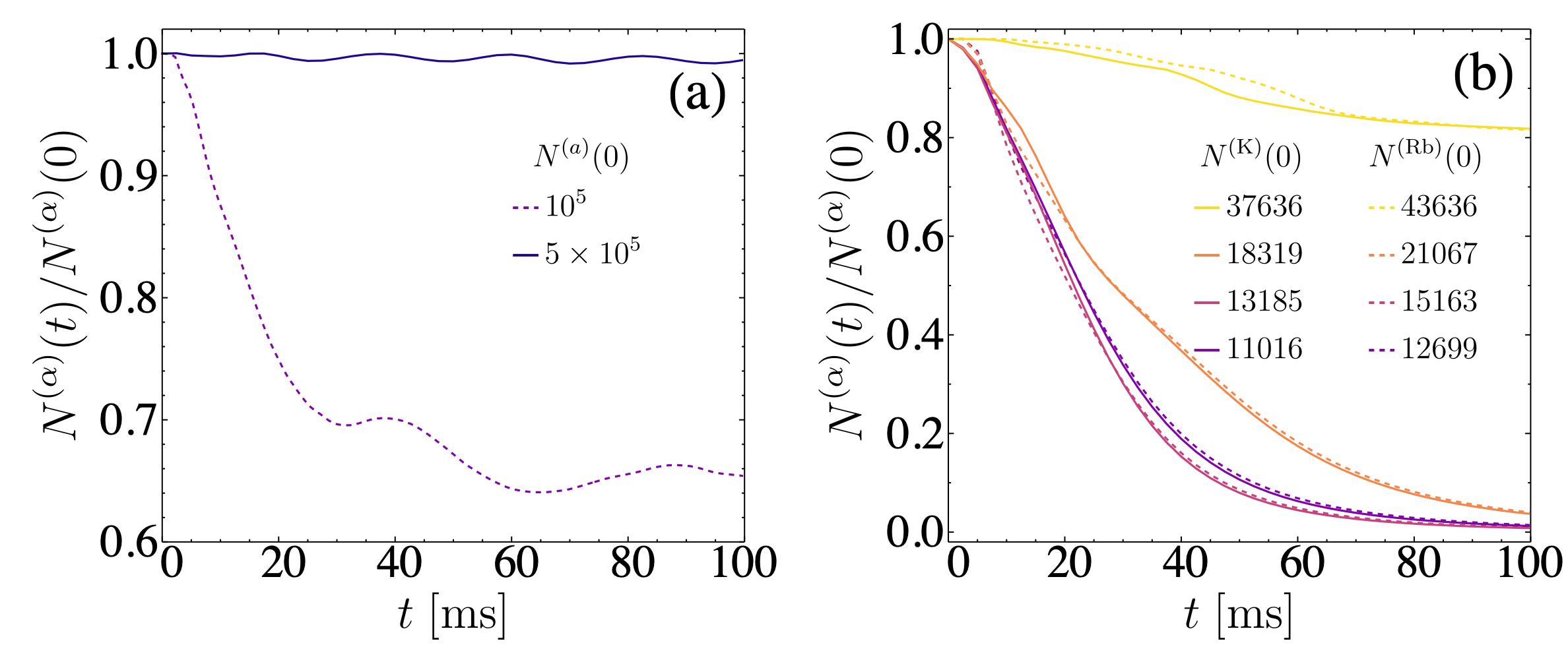}
\end{center}
\caption{Number of atoms that remain within a sphere with the minimum radius to include the complete atom cloud under a quadrupole excitation under ideal conditions, but evaluated from EGPE without losses expected from three--body scattering.  Near $N_c$ the excitation energy gives rise to enough atom losses to dissociate the droplet, in the incompressible regime such excitations occur at a better defined surface region and the self-evaporation gives rise to a lower percentage particle losses.}\label{fig:selfevaporation}
\end{figure}

In Fig.~\ref{fig:selfevaporation} the evolution of self-evaporation is illustrated for homo--nuclear, Fig.~\ref{fig:satdenhetero}a, and
hetero--nuclear, Fig.~\ref{fig:satdenhetero}b, mixtures. In the former case, $N^{(\alpha)}(t=0) =100000$  the excitation functions $u_q^{(\alpha)}$ and $v_q^{(\alpha)}$ are assumed as given by the {\it ansatz} 2 option. This number of atoms was chosen because the ground state would be stable both within the EGPE and MC formalisms. It can be observed that the self evaporation rate is not uniform in time and that an asymptotic state is reached where the quantum dropplet is stable $N^{(a)}(t\rightarrow \infty)\sim 62500$. The numerical simulations show that this state corresponds to the ground state of this number of atoms. That is, the excitation energy is released by the loss of atoms, but the final state is a self-confinement state. For greater values of 
$N^{(a)}$ less atoms are lost by evaporation, until a state is reached where excitation prevails without significant self-evaporation. This states are exemplified by the $N^{(\alpha)}(t=0) =500000$ in the Figure, the {\it ansatz} 1 was used. 

For hetero--nuclear mixtures, the self-evaporation curve exhibits  richer time  and atom number structures. First, let us focus on values of $N^{(a)}(t=0)$   for which $−\mu_K < −\mu_{Rb}$.  They are at the compressible regime and natural excitations are better described by $ansatz$ 2. At short times, more Potassium than Rubidium atoms are evaporated but only with a slightly different rate. Afterwards the  evaporation rates are even more similar until the droplet evaporates completely.
Out of the compressible regime and where the natural excitation modes are better described by $ansatz$ 1, both $^{87}$Rb and $^{41}$K evaporation rates are smaller and an asymptotic droplet ground state with fewer atoms is reached; in the intermediate time, the relation between the relative rate of $^{87}$Rb and $^{41}$K atoms release oscillates in time. It seems to be the result of a competition between the energy liberated by a particle leaving the droplet and that arising from the self-trapping of atoms with the adequate  proportion,  $N^{(a)}/N^{(b)}$.
 For larger $N^{(a)}(t=0)$ self-evaporation is suppressed, the normal mode excitation is allowed, and the atomic droplets basically behave as expected from the ideal evolution described in Section~\ref{Sec:ideal}. 

\subsection{Three--body scattering effects}

As mentioned in the Introduction, the experimental observation of the dynamics of quantum droplets before and after a collision is  highly constrained by
atom losses. For the homo--nuclear case, the biggest atom loss is due to three--body scattering with typical time scale in the 10 ms range.
Its effects can be studied by solving  the extended 
time-dependent Gross-Pitaevskii equations with the addition of imaginary terms that emulate atom losses by three--body scattering,
\begin{eqnarray}
 i \hbar \partial_t \Psi_\alpha &=&\Big( 
- \frac{\hbar^2}{2m_\alpha} \nabla^2 +   g_{\alpha \alpha} |\Psi_\alpha|^2  + g_{\alpha \beta} |\Psi_\beta|^2\nonumber\\&+&
 \frac{4}{3 \pi^2} \frac{m_\alpha^{3/5} g_{\alpha \alpha}}{ \hbar^3} (m_\alpha^{3/5} g_{\alpha \alpha} |\Psi_\alpha|^2 + m_\beta^{3/5} g_{\beta \beta} |\Psi_\beta|^2 )^{3/2} 
+iK_{\alpha\beta\gamma} \Psi_\beta^*\Psi_\gamma \Big) \Psi_\alpha. \label{EGPE_hetero--nuclear}
\end{eqnarray} 
For homo--nuclear mixtures of $^{39}$K and the general conditions under consideration, the biggest $K_{\alpha\beta\gamma}$ values correspond to collisions between atoms in the same hyperfine state  \cite{Cabrera2018,Semeghini2018,Ferioli2019}, $K_{ 111} = 6 \times 10^{− 41} m^6/ s$ and
$K_{222} = 5.4\times 10^{−39} m^6/s$. We performed numerical simulations of the evolution of  the quantum droplet that confirmed the experimental time scale. To that end, we added a confining potential which avoided the atom losses from the trap at short times. Later, this potential is turned off and
the evolution is numerically followed.

\begin{figure}[h!]
 \begin{center}
\includegraphics[width=0.75\textwidth]{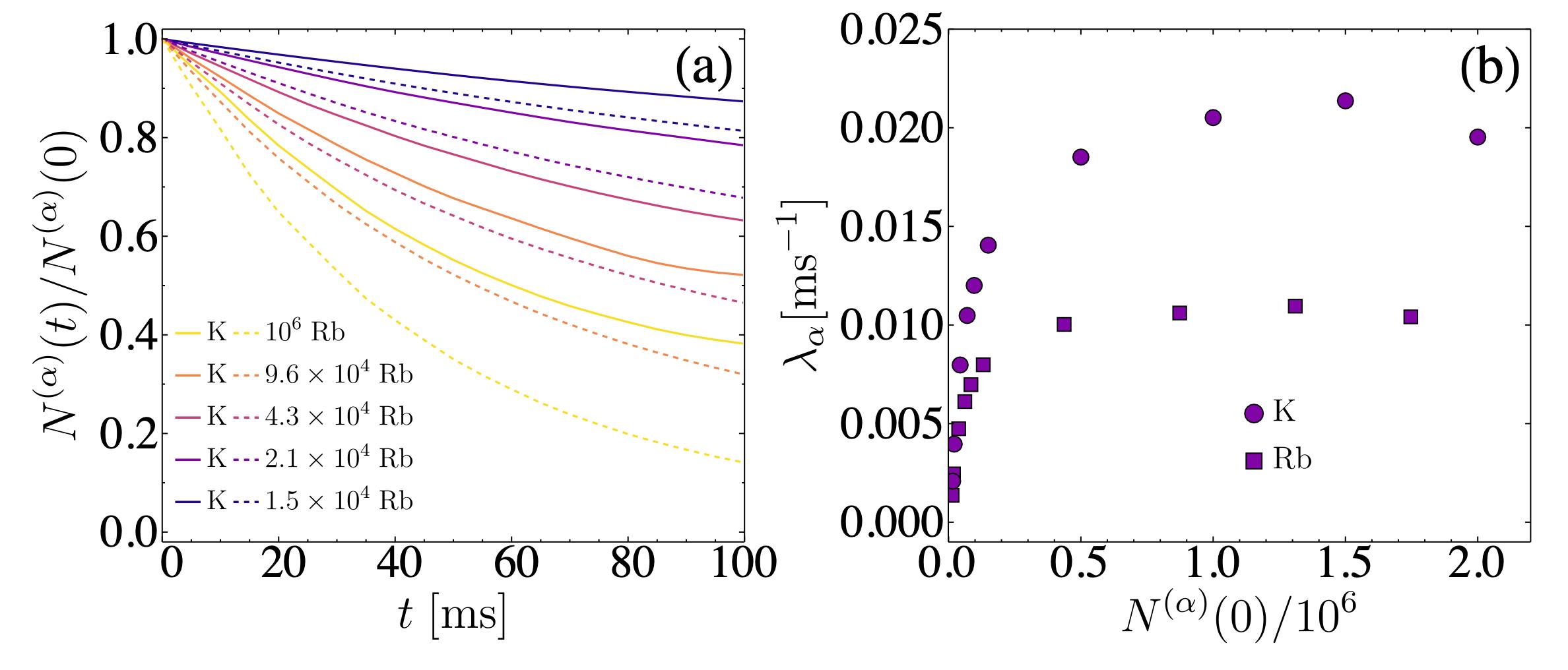}
\end{center}
\caption{Atom losses due to three--body scattering effects as described by Eq.~\ref{EGPE_hetero--nuclear} (a) Fraction of  $^{87}$Rb atoms (dashed lines) and  $^{41}$K atoms (continuos lines) that remain within a sphere with radius $R_0 + dR$ as a function of time. The initial number of $^{87}$Rb atoms is shown in the inset, the corresponding initial number of  $^{41}$K atoms satisfies the self-trapping relation at $t=0$, $N^{(K)}(0)/N^{(Rb)}(0) =\sqrt{g_{RbRb}/g_{KK}}$. (b) Behavior of the exponent $\lambda_\alpha$ that approximately describes the atoms losses shown at (a).}\label{fig:TBLab}
\end{figure}

The great advantage of using the hetero--nuclear mixtures of $^{41}$K - $^{87}$Rb instead of the two hyperfine states of $^{39}$K \cite{DErrico2019}
is that the three--body losses for the first case are  smaller compared to the second case. For the hetero--nuclear mixture  the dominant channel of losses correspond to K-Rb-Rb scattering, $K_{KRbRb} = 7 \times 10^{− 41} m^6 /s$.
We have  evaluated the time evolution of the atom losses taking as initial state the ground wave function of a droplet obtained from the EGPE. The number of atoms that remain within a sphere of radius $R_0 + dR$, with those parameters evaluated as described in Section~\ref{sec.groundhetero}, was used to monitor the atom losses. The results are illustrated in Fig.~\ref{fig:TBLab}a. An exponential fit with a parameter $\lambda_a$ that depends on the atomic species $a$ and the initial number of atoms $N_a$ were also evaluated, Fig.~\ref{fig:TBLab}b. It can be observed that, in general, losses induced by three--body scattering still involve greater rates than self-evaporation. This is also a consequence of the difference between  $^{41}$K - $^{87}$Rb which break the proportion condition that is required for self-trapping. Nevertheless there is a time window in the tens of milliseconds scale where the droplet state is preserved. For such time intervals the collision of two droplets could be observed. In fact, an experimental set-up of collisions has already been reported in the compressible regime, Ref.~\cite{Ferioli2019}.
In the following Section we describe the expectations for atom droplets in and out this regime.

\section{Frontal collisions of quantum droplets.}
The observation of the frontal collision of two quantum droplets were first reported in Ref.~\cite{Ferioli2019}.
The corresponding experiments involved homo--nuclear mixtures  of $^{39}$K atoms in two different hyperfine states.
The two droplets were generated from degenerate gases located nearby the minima of a double 
well potential. Subsequently, the potential barrier was turned off and the two droplets
migrate towards the center of a harmonic confining potential. After a controlled time interval $\Delta t$
the latter potential was also turned off, and the droplets collided in a potential free environment; $\Delta t$ 
is used to tune the initial relative speed of the droplets~\cite{Dupont2021}. If that speed were higher than a critical value $v_c$
as a result of the collision the droplets coalesced; $v_c$ depends in a non monotonic way on the number of atoms in each
initial droplet. The experiments involved $N\simeq 10^4$ so that the droplets were in the compressible regime; due to large three--body atom losses
 the whole dynamics was required to involve small time intervals of around 10 ms.

In the following paragraphs were report simulations that consider the collision of droplets that initially were radially spaced at a distance $2d_0$;
they are assumed to be kicked off as described by a plane wave wavefunction factor,

\begin{eqnarray}
\Psi(\vec r, t=0) &=&\Psi_1(\vec r) + \Psi_2(\vec r)  \nonumber\\
&=&\begin{pmatrix}\psi_{a_1}(\vec r + \vec d_0)\\ \psi_{b_1}(\vec r+ \vec d_0)\end{pmatrix} e^{i\vec k_0\cdot \vec r/2}
+\begin{pmatrix}\psi_{a_2}(\vec r - \vec d_0)\\ \psi_{b_2}(\vec r- \vec d_0\end{pmatrix} e^{-i\vec k_0\cdot \vec r/2}. \label{eq:initial}
\end{eqnarray}
The mean value 
\begin{equation}
\mathcal{K} =-\frac{\hbar^2}{2}\sum_{i=1,2}\int d^3r \Psi_i^\dagger(\vec r)\begin{pmatrix}  \frac{\nabla^2}{m_{a_i}}  & 0\\ 0 & \frac{\nabla^2}{m_{b_i}}  \end{pmatrix} \Psi_i(\vec r).\end{equation}
is taken as a measure of the initial kinetic energy of the droplets. Assuming a non significant overlap between the droplets at $t=0$,
it can be decomposed as the sum of the traslational kinetic energy of each droplet as a whole $\mathcal{K}_{tras}$, and an  internal kinetic energy of the atoms within the droplets $\mathcal{K}_{int}$
\begin{eqnarray}
&\approx& \mathcal{K}_{trans} + \mathcal{K}_{int}\label{eq:trasplusint}\\
\mathcal{K}_{tras}&=& \frac{\hbar^2}{2}\Big[ \frac{N^{(a_1)}k_0^2}{4m_{a_1}} + \frac{N^{(b_1)}k_0^2}{4m_{b_1}} +
\frac{N^{(a_2)}k_0^2}{4m_{a_2}} + \frac{N^{(b_2)}k_0^2}{4m_{b_2}}  \Big]\\
 \mathcal{K}_{int} &=& -\frac{\hbar^2}{2}\sum_{i=1,2}\sum_{\alpha=a,b}\int d^3r\psi^*_{\alpha_i}(\vec r)\frac{\nabla^2}{m_{\alpha_i}}\psi_{\alpha_i}(\vec r)
\end{eqnarray}
The dimensionless quantity suitable for characterizing binary collisions of
classical incompressible droplets is the Weber number \cite{frohn2000}. It is a  dimesionless
quantity  for each one of the droplets, defined as the ratio between the traslational kinetic energy $\mathcal{K}_{tras}$ before the collision and the surface energy of excitation evaluated in terms of the surface tension of the droplet. This definition can be extended to quantum droplets as
\begin{equation}
\mathrm{We}_\ell =\frac{\mathcal{K}_{tras}}{R_0^2\sigma_\ell}\label{eq:Weber}
\end{equation}
where $R_0$ is the initial radius of the droplet. 
For the collision of two identical quantum droplets like those described in this work, this expression becomes 

{\bf Ansatz 1}
\begin{eqnarray}
\mathrm{We}_\ell^{(Ans1)}&=& -\frac{\pi k_0^2}{3 R_0^2} \frac{ \left( \int dr \partial_r \rho^{(a)}_0 r^{\ell+2} \right)^2 } { \int dr  (\partial_r \phi^{(a)}_0)^2  
\int dr   \partial_r \rho^{(a)}_0 r^{2\ell+1} 
  }  \\
&=& \frac{8\pi (dR k_0)^2}{3}\Big(\frac{dR}{R_0}\Big)^2 \frac{((\ell +2)!)^2}{(2\ell +1)!}\Big[1-\frac{1}{(1+ e^{R_0/dR})^2}\Big]^{-1}
                      \Big[\frac{(F_{\ell+1}(R_0/dR))^2}{F_{2\ell }(R_0/dR)}\Big], \label{eq:We1}
\end{eqnarray}
{\bf Ansatz 2}
\begin{eqnarray}
\mathrm{We}_\ell^{(Ans2)}&=& \frac{\pi k_0^2}{3 R_0^2}\frac{  \int dr  \, \partial_r \rho^{(a)}_0 r^{2\ell+1} }{ \int dr  \,  (\partial_r \phi^{(a)}_0 )^2  r^{2\ell-2}   }
  \\
&=&\frac{16\pi (dR k_0)^2}{3}\Big(\frac{dR}{R_0}\Big)^2 (\ell(4\ell^2 -1))
                      \frac{F_{2\ell}(R_0/dR)}{F_{2\ell-3}(R_0/dR)+F_{2\ell -4}(R_0/dR)}.\label{eq:We2}
\end{eqnarray}

Notice that the expressions of $\mathrm{We}_\ell$ for the ground state of the quantum droplets given by Eqs.~(\ref{eq:We1},\ref{eq:We2}) make evident the relevance of 
the relation between the de Broglie wavelength and the skin width of the droplet as given by $k_0 dR$ as a measure of the momentum that could be transfered to the droplet during the collision. The second relevant parameter is the ratio $R_0/dR$ that determines the deformation of the droplet due to a multipole excitation. 

\subsection{Frontal collisions of two droplets with and without out atom losses by three--body scattering.}
At first, we have  studied the effects of binary collisions of quantum droplets in an idealized scheme described by  Eqs.~(\ref{EGPE_hetero--nuclear}) with  $K_{\alpha\beta\gamma} = 0$. We considered both homo--nuclear droplets --$^{39}$K, with $\mathrm{a}_{aa} = 48.57\mathrm{a}_0$ and $\mathrm{a}_{ab} = 51.36\mathrm{a}_0$ --
and hetero--nuclear droplets ---$^{39}$K and $^{87}$Rb with $\mathrm{a}_{KK} =62\mathrm{a}_0$ $\mathrm{a}_{RbRb} = 100.4\mathrm{a}_0$, and $\mathrm{a}_{KRb} =-82.0\mathrm{a}_0$~--. Different initial kinetic energies and number of particles were studied always in the incompressible regime where no autoevaporation is expected.  The initial condition corresponds to identical ground state droplets with an overall movement described by Eq.~(\ref{eq:initial}). Later on, simulations including the adequate scattering coefficients $K_{\alpha\beta\gamma} \neq 0$ were performed.

\begin{figure}[ht!]
 \begin{center}
\includegraphics[width=0.6\textwidth]{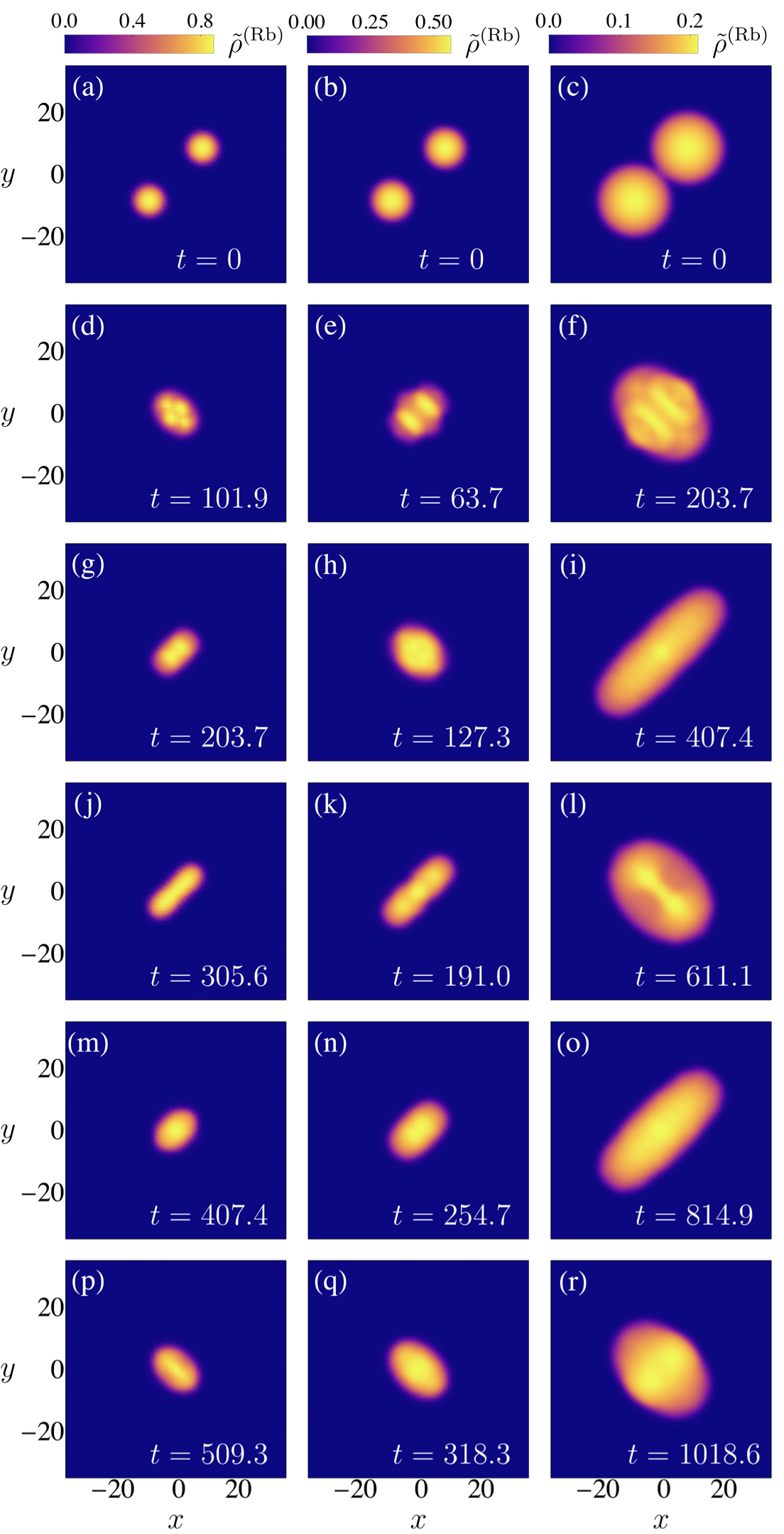}
\end{center}
\caption{Illustrative examples of frontal collisions of two droplets leading to its coalescence. Identical hetero--nuclear droplets of Rb and K atoms are considered in each column, where the density of Rb atoms integrated along the z-direction $\tilde \rho^{(Rb)}(x,y) = 
\int dz  \rho^{(Rb)}(x,y,z)$ is depicted at different evolution times. The initial translational kinetic energy is varied to guarantee that for  
$N^{Rb} = 5 \times 10^5$ (first column),  $N^{Rb} = 1 \times 10^6$ (second column) and $N^{Rb} = 5 \times 10^6$ the Weber number equals $\mathrm{We}_2^{(ans1)}= 6.2  $. }\label{fig:coalescence}
\end{figure}

\begin{figure}[h!]
\begin{center}
\includegraphics[width=0.6\textwidth]{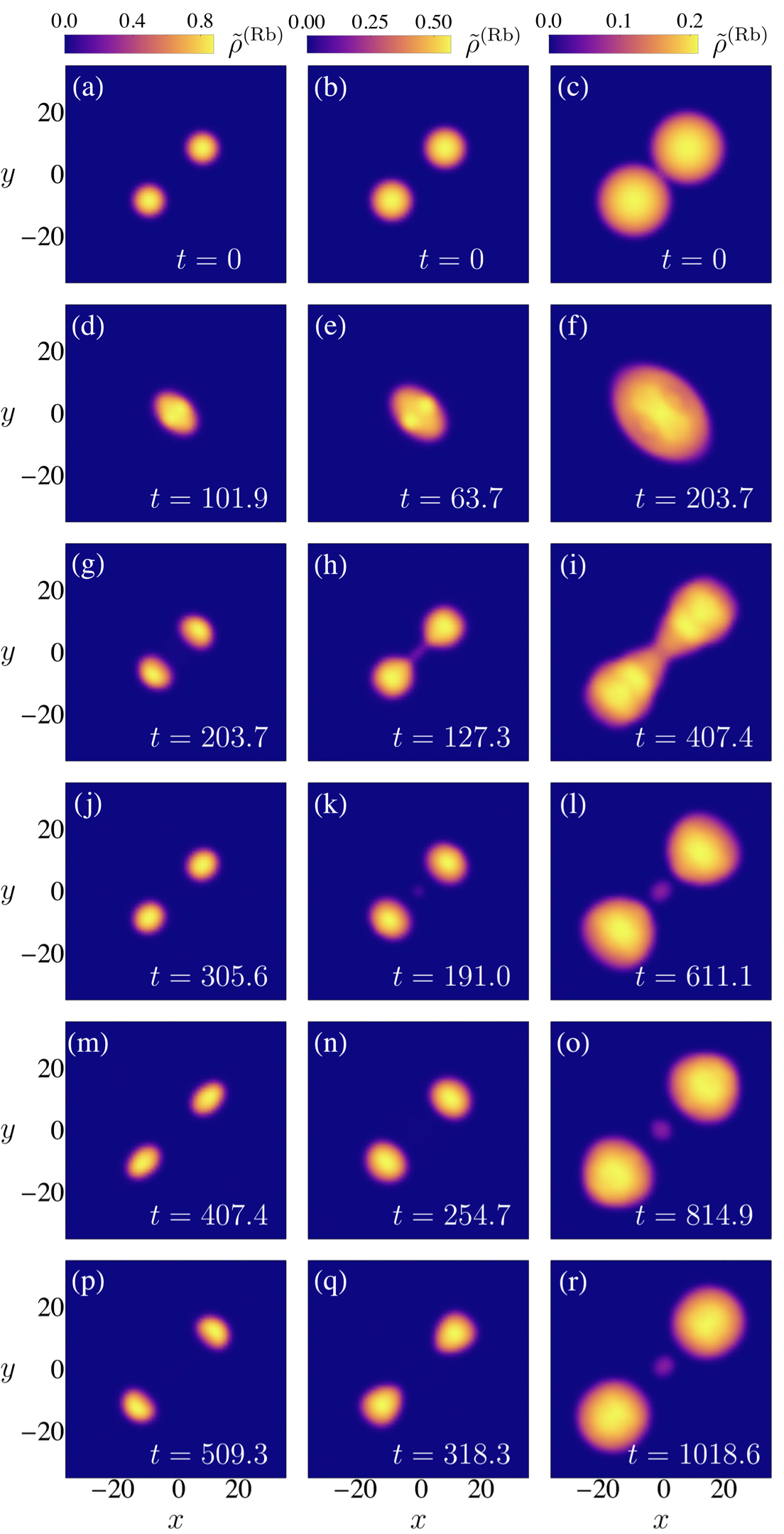}
\end{center}
\caption{Illustrative examples of frontal collisions of two droplets leading to an initial coalescence followed by a
disintegration of the resulting droplet into two droplets that are then expelled in opposite directions. Initially,
identical hetero--nuclear droplets of Rb and K atoms are considered in each column, where the density of Rb atoms integrated along the z-direction $  \tilde{\rho}^{(Rb)} (x, y) = \int dz \rho^{(Rb)} (x, y, z)$ is depicted at different evolution times. The initial
translational kinetic energy is varied to guarantee that for $N^{(Rb)} = 5 \times 10^5$ (first column), $N^{(Rb)} = 1 \times 10^6$ (second column) and
$N^{(Rb)} = 5 \times 10^6$ (third column), the Weber number equals  We$^{(ans1)}_2 =14.5$.}
\label{fig:breaking}
\end{figure}

\begin{figure}[h!]
 \begin{center}
\includegraphics[width=0.6\textwidth]{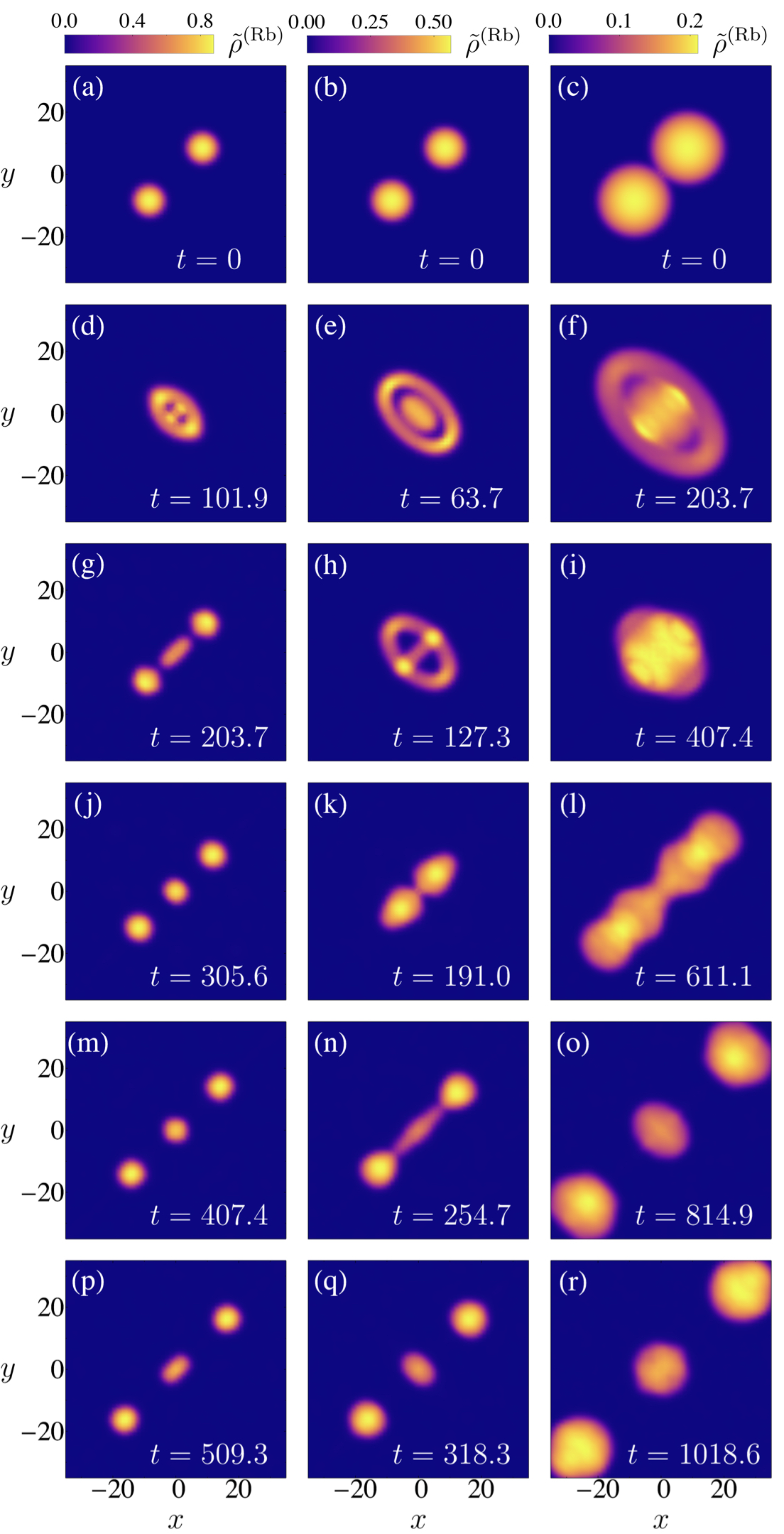}
\end{center}
\caption{Illustrative examples of frontal collisions of two droplets leading to an initial coalescence followed by a disintegration of the resulting droplet into two droplets that are then expelled in opposite directions and a third droplet is formed at the center.. Initially, identical hetero--nuclear droplets of Rb and K atoms are considered in each column, where the density of Rb atoms integrated along the z-direction $\tilde \rho^{(Rb)}(x,y) = \int dz 
 \rho^{(Rb)}(x,y,z)$ is depicted at different evolution times. The initial translational kinetic energy is varied to guarantee that for  $N^{Rb} = 5 \times 10^5$ (first column),  $N^{Rb} = 1 \times 10^6$ (second column)
and $N^{Rb} = 5 \times 10^6$ the Weber number equals $\mathrm{We}_2^{(ans1)}= 31.9$.}
\label{fig:coalescence-breaking}
\end{figure}

In general, for the kinetic energies here considered, the simulations yield different possible final results of a frontal collision,
\begin{itemize}
\item[(i)] Coalescence: Formation of a single droplet from two colliding droplets. The resulting droplet is found to be created mainly in a
quadrupole excitation, although higher order modes could also be present (illustrated in Fig.~\ref{fig:coalescence}).

\item[(ii)] Disintegration into two droplets after coalescence: after the collision, the oscillations of the resultant coalesced droplet are so strong that it breaks into two similar droplets. These two droplets exit in opposite directions and leave the collision zone with a translational movement described mainly as a dipole like oscillation (illustrated in Fig.~\ref{fig:breaking}).

\item[(iii)]  Disintegration into three droplets after coalescence: when the initial droplets are large and the collision is energetic enough,
after the breaking and the expulsion of two droplets in opposite directions, a third droplet is formed at the center. The latter oscillates mainly in a quadrupole mode as in the first case (illustrated in Fig.~\ref{fig:coalescence-breaking}). Disintegration into more droplets is expected for even larger traslational kinetic energies.
\end{itemize}
 Similar results for homo-nuclear droplets are described in Ref.~\cite{Cikojevic2021}.

\begin{figure}[ht!]
 \begin{center}
\includegraphics[width=0.75\textwidth]{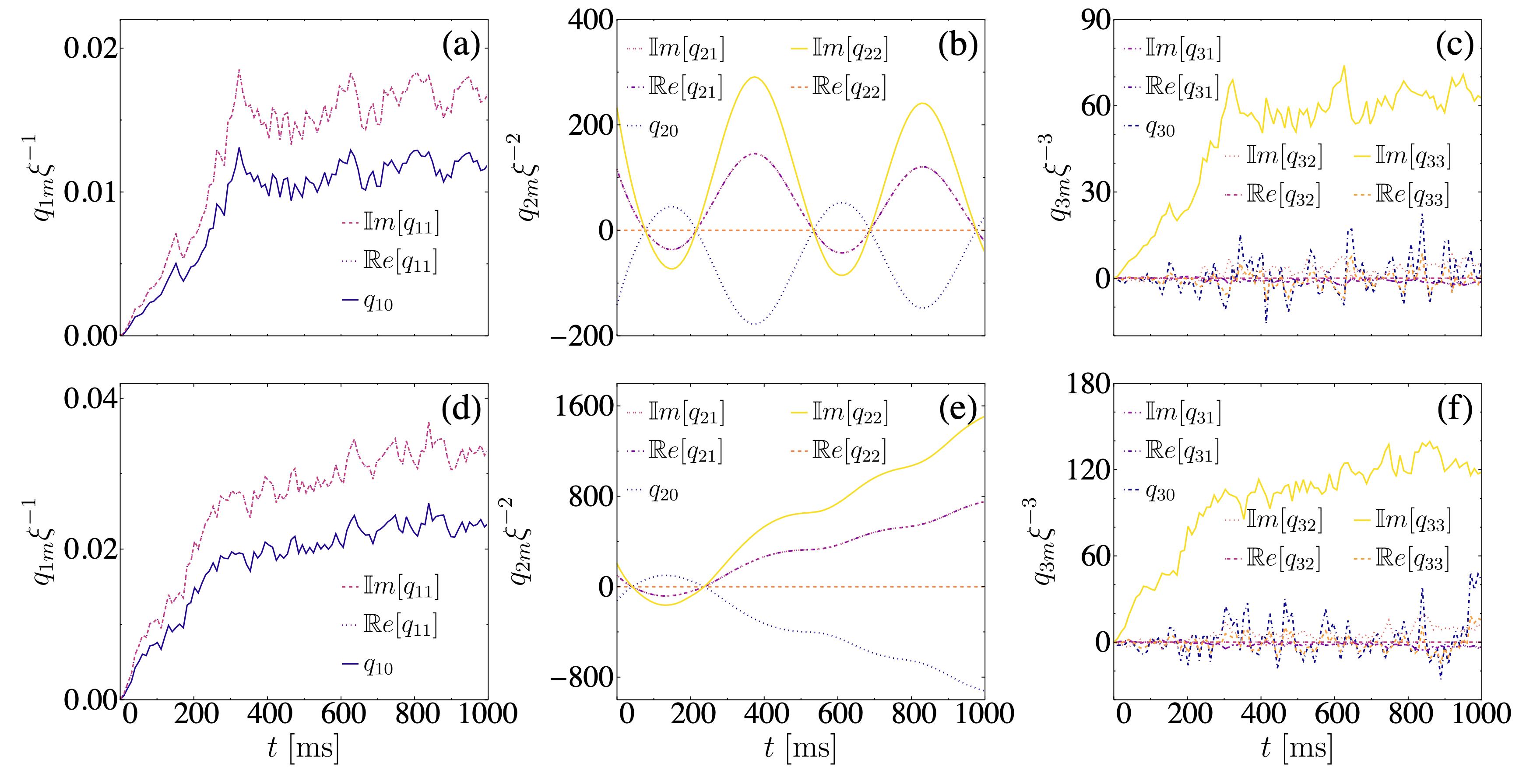}
\end{center}
\caption{Illustrative examples of multipole moments as a function of time that result from the ideal (without three--body losses) frontal collision of two identical hetero--nuclear Rb-K droplets. The first row corresponds
to a collision between droplets with $N^{(Rb)} = 5\times 10^6$ and $\mathrm{We}_2 = 5.87 $ that culminates with the coalescence of the droplets. The second row corresponds
to $N^{(Rb)} = 5\times 10^6$ and $\mathrm{We}_2 =16.8 $ that culminates with a disintegration into two droplets. (a,d)  dipole moment (b,e) quadrupole moment and (c,f) the octupole moment. }\label{fig:multipole}
\end{figure}

Now, we can explore the multipole character of the collective excitations of the droplets as characterized in terms of the evolution over time of the multipole moment of the density of the atoms,
\begin{equation}
q_{\ell m} (t) = \int d^3x_0 Y_{\ell m}(\theta_0,\varphi_0)r_0^\ell \vert\psi_a(x_0,t)\vert^2,
\end{equation}
the quantization axis is taken as the collision axis $\hat z = (1, 1, 1)$. In Fig.\ref{fig:multipole} the results are illustrated for the same examples depicted in Figs.~\ref{fig:coalescence}-\ref{fig:coalescence-breaking}. 
The quadrupole moments oscillate around zero, the first crossing $q_{2m} = 0$ coincides with the time at which the centers of mass of the individual droplets coincide. The second crossing $q_{2m} = 0$ signals the moment at which a maximum amplitude of oscillation takes place. If the final result of the collision corresponds to a breaking of the droplet, it is at this time when the quadrupole moment begins to grow and finally diverges (see Fig.~\ref{fig:multipole}).

\begin{figure}[h!]
 \begin{center}
\includegraphics[width=0.75\textwidth]{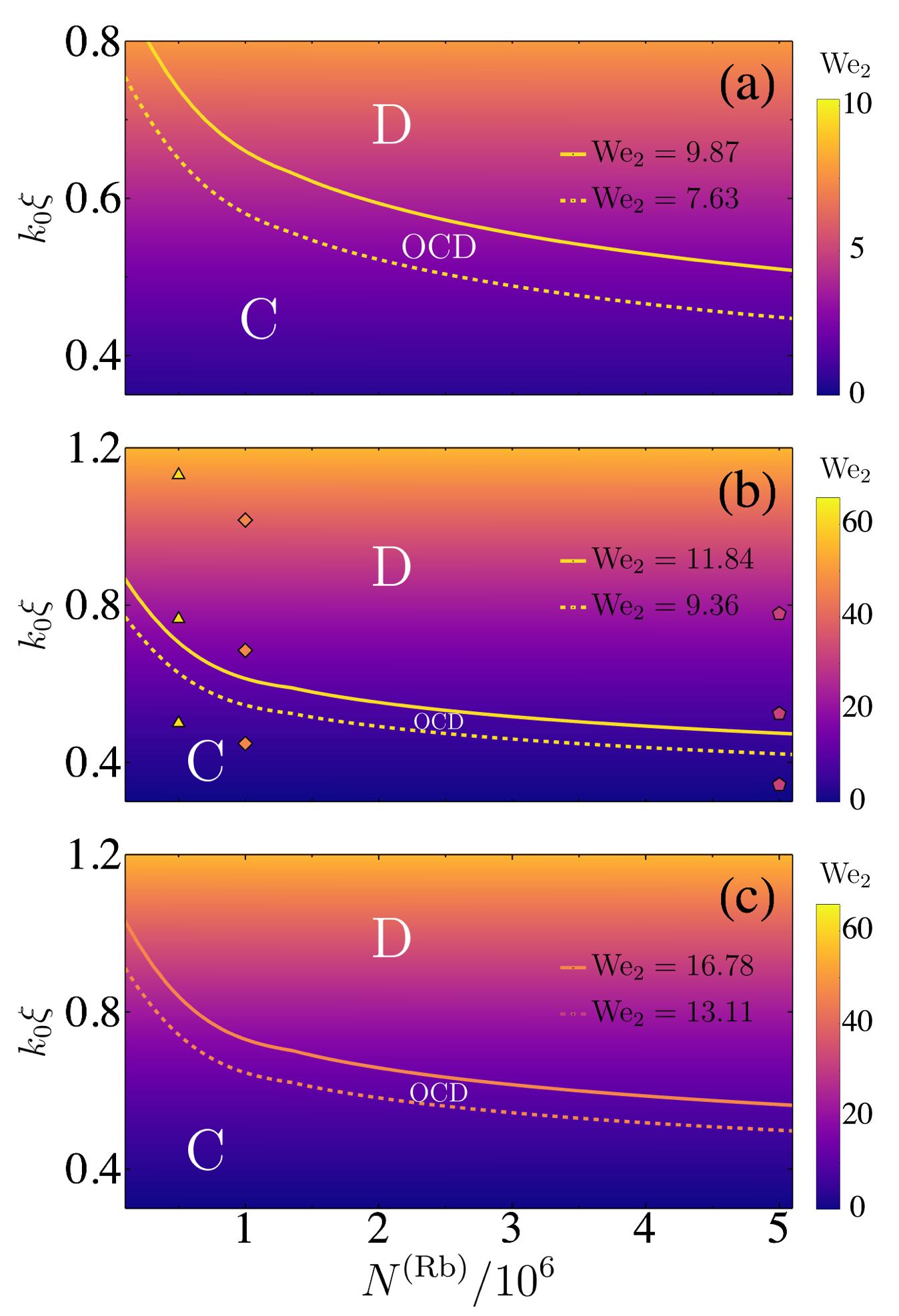}
\end{center}
\caption{Quadrupole Weber number $\mathrm{We}_2$ as a function of the number of atoms of a given species is illustrated for (a) homunuclear and (b)
hetero--nuclear droplets (c) hetero--nucleat droplets including atom losses. In the first case N corresponds to half the total number of atoms in the droplet and in case
(b) and (c) to $N^{(Rb)}$ in an Rb and K admixture. The outcome of a frontal collision will give rise to the coalescence of the
droplets for $\mathrm{We}_2$ below the lower line (C region).
The disintegration of the originally droplet formed during the collision  is observed for
$\mathrm{We}_2$ above the higher line (D region). For values of $\mathrm{We}_2$ between these lines involve a long transient stage where the coalesced
droplet oscillates strongly before its disintegration (OCD region). The dots in (b) correspond to the values of $k_0\xi$ and $N^{(Rb)}$ illustrated in previous Figures.}\label{fig:Wecol}
\end{figure}

The Weber number given by Eq.~(\ref{eq:Weber}) must be evaluated according to the expression of the surface tension adequate for the regime at which the initial droplets are prepared: in the incompressible regime {\it ansatz} 1 is used while in the compressible regime {\it ansatz 2} is the correct one.    Figure \ref{fig:Wecol}, illustrates  the Weber number as a function of the number of atoms and the wave vector $k_0$. 
When three--body losses are included, the Weber number is still a good parameter to delimit the
immediate breakdown of a droplet after the collision. However the transient regime is linked to a different band of $\mathrm{We}_2$ values.
A rich variety of transient configurations that culminate in one of the three above results are found. Their observation depends highly on the inclusion of three--body atom losses. In the ideal case $K_{\alpha\beta\gamma} = 0$, frequently after the break into two droplets like that described in (ii), the dipole like oscillation of each individual droplet is faster than their relative translation and posteriori coalescence occurs--described by (i), the final single droplet remains oscillating mainly in a quadrupole way. The transient regime usually takes place for quadrupole  Weber numbers between two given values $\mathrm{We}_\ell^{min}$ and $\mathrm{We}_\ell^{max}$ whose values are exemplified in Fig.~{\ref{fig:Wecol}} for collisions between homo-nuclear and hetero-nuclear droplets. For $\mathrm{We}_\ell < \mathrm{We}_\ell^{min}$ the scenario described by (i) takes place without a transient stage. For $\mathrm{We}_\ell > \mathrm{We}_\ell^{max}$ the scenario described by either (ii) or (iii) takes place without a transient stage. Figure~{\ref{fig:Wecol}} is supported by extensive numerical simulations carried out for the values of the parameters $k_0\xi$ and $N$ reported in it.

When three--body losses are included, the observation of the results of the collision in terms of quantum droplets is highly dependent on the initial conditions.
For a given $\vec k_0$, the parameter $\vec d_0$ should be chosen so that atom losses by three--body scattering are small-- less than $2\%$ before the collision takes place-- and, simultaneously, the droplets overlap at $t = 0$ is avoided. The mean life time of the hetero--nuclear droplets is in the range of tens of milliseconds as expected from Fig~(\ref{fig:TBLab}) and confirmed by experimental evidence \cite{DErrico2019}. So that, the long time evolutions like those described in the ideal case are difficult to observe for low $k_0$ values.
Nevertheless, in spite of the short times involved, the regimes at which the scenarios (i)-(iii) occur were observed in the simulations. An interesting effect due to  the atomic gas that surrounds each one of the droplets is also predicted: the  surrounding evaporated gas created nearby a given droplet interact with the atoms from the other droplet through contact terms. This can result in a reconfiguration of a merged quantum droplet. The effect is similar to that exemplified by Fig.6c-d.
Taking into account three--body losses, the Weber number is still a good parameter to delimit the
immediate breakdown of a coalesced droplet after the collision.

\section{Discussion.}
In this work we have studied theoretically the static and dynamical properties of quantum droplets constituted by binary mixtures of homo-nuclear and hetero--nuclear ultracold atoms. One of the more interesting features of quantum droplets corresponds to self-trapping.
Our calculations show that the effective equations as those built as extensions to the Gross-Pitaevskii formalism provide robust interpretations of the droplets behavior
both in the compressible and the incompressible regime. Our calculations confirm the slight differences between the expectations derived  from Monte Carlo studies  and the LHY-EGPE. They manifest not only on the minimum number of atoms required for self-trapping, but also on the radius $R_0$ and thickness of the surface $dR$ of the droplets; their dependence on the scattering lenghts is discussed at depth in Ref.~\cite{Cikojevic2018,Boronat2021}

We have also found that in the incompressible and compressible regimes the elementary Bogolubov excitations, Eq.~(\ref{eq:ansatz}-\ref{eq:srI}), are better described in a variational basis by Eqs.~(\ref{eq:a1}) and (\ref{eq:a2}) respectively.  The spherical symmetry of the droplets is directly incorporated in the selection of the basis. The corresponding expressions for the surface tension reflect such a symmetry, and  exhibit a different behavior on the multipole order. In the incompressible regime this dependence coincides for {\it ansatz 1} with that expected for classical liquids and nuclear matter. The excitation modes and accompaining surface tension derived from {\it ansatz 2} are more adequately described for a fluid in the compressible regime. For homo-nuclear mixtures, differences in the predicted shape between the effective range formulation and the EGPE
manifest also on slight differences in the excitation energies of the Bogolubov modes. 

Self-evaporation is a particularly interesting phenomenon predicted for quantum droplets that is not
easily observed due to its competition with atom losses  arising from three--body scattering. We have identified two other relevant factors that accelerate the disintegration of the quantum droplet: loss of the adequate proportion of each atomic species and differences in spatial configuration of each species.  Phenomena that could be more easily observed for hetero-nuclear droplets.

In spite of the difficulties to experimentally study the dynamics of quantum droplets, we have theoretically shown that if the droplets are excited via frontal binary collisions it is still possible to observe a non trivial dynamics of these exotic droplets. The Weber number resulting from the surface tension expressions introduced in our study exhibit a simple dependence on the radii of the droplets $R_0$ and the relative droplets momentum $\hbar k_0$ both of them measured in terms of the surface width $dR$, Eqs.~(\ref{eq:We1},\ref{eq:We2}) which makes explicit the surface character of the acompaining excitations.   We have also shown that the
Weber number evaluated with the adequate {\it ansatz} define different regimes of excitation that
include conditions for (i)  the formation of pairs of quantum droplets excited as oscillating dipoles that detach from each other leaving the impact region, (ii) the formation of a single droplet (mainly in  a quadrupole excitation mode) that merges from two elementary droplets; (iii) if three--body losses are included space regions are created where the lost atoms interact via de contact terms and lead to a reconfiguration of otherwise disjoint droplets (this effect is similar to that exemplified by Fig.6c-d but it is enhanced by the evaporated atoms).  Contrary to analysis that evaluate the surface tension of the planar interface
described by local energy functionals and latter on incorporate curvature through Tolman like terms \cite{Tolman1949}, no extra parameters are required within our formalism to identify the different collision regimes.

We hope that these theoretical analyses pave the way to further studies of novel states induced in ultracold atoms and their analog to other quatum matter systems.

\section{Acknowledgements }

This work was partially supported by the grants DGAPA-PAPIIT, UNAM: IN-103020, IN-109619, UNAM-AG810720,  and CONACYT Ciencia B\'asica: A1-S-30934 and Laboratorios Nacionales 315838. E. Alba-Arroyo thanks CONACyT for his posgraduate studies fellowship  464666 and DGAPA-PAPIIT IN-103020 graduation support.

\end{document}